\documentclass[default]{aastex63}

\newcommand\APOLLO{\texttt{APOLLO}}
\newcommand\JWST{\textit{JWST}}
\newcommand\JHK{$JHK_\mathrm{s}$}

\usepackage{amsmath}
\usepackage{amssymb}
\usepackage{rotating}
\usepackage{multirow}

\received{2022 June 14}
\revised{2023 July 30}
\accepted{2023 September 18}
\submitjournal{AJ}

\shorttitle{Atmospheric Retrieval of HD 106906 b}
\shortauthors{Adams et.~al.}

\graphicspath{{./}{./}}

\begin{document}

\title{Atmospheric Retrieval of L Dwarfs: Benchmarking Results and Characterizing the Young Planetary Mass Companion HD 106906 b in the Near-Infrared}

\correspondingauthor{Arthur D. Adams}
\email{arthura@ucr.edu}
\author[0000-0002-7139-3695]{Arthur D. Adams}
\affiliation{Department of Earth and Planetary Sciences, University of California, Riverside, CA 92521, USA}
\affiliation{Department of Astronomy, University of Michigan, 1085 S. University, Ann Arbor, MI 48109, USA}
\author[0000-0003-3963-9672]{Michael R. Meyer}
\affiliation{Department of Astronomy, University of Michigan, 1085 S. University, Ann Arbor, MI 48109, USA}
\author[0000-0002-4884-7150]{Alex R. Howe}
\affiliation{NASA Goddard Space Flight Center, 8800 Greenbelt Rd, Greenbelt, MD 20771, USA}
\author[0000-0003-4600-5627]{Ben Burningham}
\affiliation{Centre for Astrophysics Research, Department of Physics, Astronomy and Mathematics, University of Hertfordshire, Hatfield AL10 9AB}
\author[0000-0001-9915-2132]{Sebastian Daemgen}
\affiliation{ETH Zürich, Institute for Particle Physics and Astrophysics, Wolfgang-Pauli-Strasse 27, 8093 Zürich, Switzerland}
\author[0000-0002-9843-4354]{Jonathan Fortney}
\affiliation{Department of Astronomy \& Astrophysics, University of California, Santa Cruz, CA 95064, USA}
\author[0000-0002-2338-476X]{Mike Line}
\affiliation{School of Earth and Space Exploration, Arizona State University, Tempe, AZ 85287}
\author[0000-0002-5251-2943]{Mark Marley}
\affiliation{Lunar \& Planetary Laboratory, University of Arizona, Tucson, AZ 85721}
\author[0000-0003-3829-7412]{Sascha P. Quanz}
\affiliation{ETH Zürich, Institute for Particle Physics and Astrophysics, Wolfgang-Pauli-Strasse 27, 8093 Zürich, Switzerland}
\author[0000-0002-9276-8118]{Kamen Todorov}
\affiliation{Anton Pannekoek Institute for Astronomy, University of Amsterdam, Science Park 904, 1098 XH Amsterdam, The Netherlands}

\begin{abstract}
We present model constraints on the atmospheric structure of HD 106906 b, a planetary-mass companion orbiting at a $\sim$700 AU projected separation around a 15 Myr-old stellar binary, using the APOLLO retrieval code on spectral data spanning 1.1–2.5 $\mu$m. C/O ratios can provide evidence for companion formation pathways, as such pathways are ambiguous both at wide separations and at star-to-companion mass ratios in the overlap between the distributions of planets and brown dwarfs. We benchmark our code against an existing retrieval of the field L dwarf 2M2224-0158, returning a C/O ratio consistent with previous fits to the same $JHK_\mathrm{s}$ data, but disagreeing in the thermal structure, cloud properties, and atmospheric scale height. For HD 106906 b, we retrieve C/O $=0.53^{+0.15}_{-0.25}$, consistent with the C/O ratios expected for HD 106906’s stellar association and therefore consistent with a stellar-like formation for the companion. We find abundances of H$_2$O and CO near chemical equilibrium values for a solar metallicity, but a surface gravity lower than expected, as well as a thermal profile with sharp transitions in the temperature gradient. Despite high signal-to-noise and spectral resolution, more accurate constraints necessitate data across a broader wavelength range. This work serves as preparation for subsequent retrievals in the era of \JWST{}, as \JWST{}’s spectral range provides a promising opportunity to resolve difficulties in fitting low-gravity L dwarfs, and also underscores the need for simultaneous comparative retrievals on L dwarf companions with multiple retrieval codes.
\end{abstract}


\section{Introduction} \label{sec:intro}
Mass is typically taken as the discriminator between planets and brown dwarfs, based on the minimum of $\sim$13 Jupiter masses needed for sustained deuterium fusion \citep{Spiegel2011}. While one can use mass alone to define the classes of planets and brown dwarfs, there is an alternate definition based on the formation pathway of an object as more ``star''-like or ``planet''-like \citep[see e.g.][]{Janson2012,Pepe2014,Currie2014,Currie2020,Schlaufman2018}. These definitions may produce similar categories of planetary and brown dwarf companions as with the mass definition. However, 13 $M_\mathrm{J}$ is not known to be a strict upper limit to forming companions as planets (i.e.~that the companion forms within a circumstellar disk surrounding a young star) --- nor is 13 $M_\mathrm{J}$ a strict minimum below which objects may not collapse from a molecular cloud. In exoplanet and brown dwarf demographics, there is a local minimum in the observed distributions of companions' masses as ratios to their hosts' masses, as seen in radial velocity and astrometry \citep{Sahlmann2010}, direct imaging \citep{Reggiani2016,Vigan2017,Nielsen2019}, and microlensing \citep{Shvartzvald2016,Suzuki2016}. That is, using the mass definition of planets and brown dwarfs, it is both difficult to form planets at masses as large as $\sim$1\% of their hosts, and also difficult to form brown dwarfs with mass ratios that small. It is companions in this region that serve as the more ambiguous cases when using formation history as the criterion for distinguishing planets and brown dwarfs. How can we tell the formation pathway for individual companions?

The chemical composition of a companion reflects its formation pathway, especially in the carbon-to-oxygen (C/O) ratio of its envelope relative to those measured in its host star\footnote{The metallicity can also provide important evidence of planet-like formation, especially for core accretion. See for example the review in \citet{Madhusudhan2019}.}. Planetary formation pathways are themselves generally divided into core accretion versus gravitational instability \citep[see e.g.][]{Forgan2013,Forgan2015,Forgan2018,Kratter2016}. Core accretion allows a planet's C/O ratio to diverge from its host's C/O based on where and when companions accrete their material in the disk. The H$_2$O, CO, and CO$_2$ ice lines determine the relative fraction of C and O contained in gases versus solids as a function of distance from the host star \citep[e.g.][]{Mousis2009,Oberg2011,Madhusudhan2011a}. Since disk chemistry also evolves in time, planet compositions will reflect the chemical evolution in the disk over their development \citep[e.g.][]{Booth2017, Madhusudhan2017}, both in the disk midplane \citep{Eistrup2016,Eistrup2018} and vertically \citep{Cridland2020b}. For planets formed via gravitational instability the formation mechanism is different, but there is still ample opportunity for the atmosphere to evolve its chemistry away from a stellar-like C/O ratio \citep[in this case, because the protoplanetary fragment has time to stratify its C and O compounds between the core and envelope; see e.g.][]{Ilee2017}.

While a non-stellar C/O ratio will certainly be reflected in the companion's observable emission spectrum, the presence of clouds in the companion photosphere require a careful modeling approach. Many young ($lesssim 100$ Myr) companions in the target mass ratio range fall in the L and T spectral types \citep{Kirkpatrick2005}. In the warmer L dwarfs (1300 K$\lesssim T_\mathrm{eff} \lesssim$2000 K), a variety of cloud species become important opacity sources \citep[see e.g.][]{Morley2012,Marley2013,Helling2014,Helling2020}; silicates such as enstatite (MgSiO$_3$) and forsterite (Mg$_2$SiO$_4$), iron (Fe), aluminum oxides (e.g.~Al$_2$O$_3$), and quartz (SiO$_2$) can all contribute significantly in column density to L dwarf photospheres \citep{Helling2006,Gao2020,Woitke2020,Burningham2021}. One important limitation in using observed gas abundances alone to constrain a C/O ratio is that oxygen-rich cloud species can condense out a significant amount of the oxygen budget at the pressures they reside. This biases the gas-derived C/O constrained from an emission spectrum, as it will be carbon-rich relative to the cumulative envelope C/O of the companion at the time of formation. We discuss our results in the context of this assumption in the discussion (\S \ref{sec:discussion}).

Fitting brown dwarf spectra has traditionally relied on interpolations using grids of forward models that rely on specific input physics \citep[e.g.][]{Burrows1993b,Allard1996,Marley1996,Tsuji1996,Saumon2000,Geballe2001,Hubeny2007,Saumon2008,Yamamura2010,Patience2012}. There have been numerous analyses of field brown dwarfs that use such libraries of model spectra to constrain global properties such as effective temperature, metallicity, age, surface gravity, luminosity, mass, and in some cases cloud layers \citep[e.g.][]{Allers2007,Cushing2008,Cruz2009,Stephens2009,Rice2010,Allers2013,Bonnefoy2014,Martin2017}. The link between theory and observation for sub-stellar atmospheres has been evolving for quite some time, as reviewed in works such as \citet{Burrows2001,Marley2013,Marley2015}, which has motivated a second approach to spectral fitting, namely atmospheric retrieval.\footnote{For a recent review of atmospheric retrieval methods, see e.g.~\citet{Fortney2021}.} Retrievals opt to generate forward models in parallel with a parameter estimation technique such as Markov-Chain Monte Carlo (MCMC) or Nested Sampling, rather than interpolate from a pre-computed grid. Such an approach is computationally intensive and typically requires one to make simplifying assumptions in the parametrization that may or may not be physically consistent. However, the potential benefit is a more direct and precise constraint on key physical parameters, which may be warranted if the spectra are sensitive to small changes in the parameters, such as those of the temperature-pressure (T-P) profile, molecular abundances, or significant cloud opacities.

We now have the results of a growing number of L dwarf retrievals to guide us in interpreting our data. \citet{Burningham2017} retrieve atmospheric properties from the near-infrared spectra of 2 mid-L field dwarfs using the \texttt{Brewster} retrieval code. This is then expanded into the mid-infrared in \citet{Burningham2021}, constraining multiple cloud species including enstatite (MgSiO$_3$), quartz (SiO$_2$), and iron (Fe). \citet{Gonzales2020} apply \texttt{Brewster} to a L+T sub-dwarf binary and provide evidence for their co-formation as well as evidence for clouds in the primary. \citet{Peretti2019} use the retrieval code \texttt{HELIOS-R} on a combination of thermal infrared photometry and $R \sim 30$ $J$-band data and place their retrieved chemical composition in context with both astrometric and radial velocity measurements. \citet{Molliere2020} employ the \texttt{petitRADTRANS} code to fit the near-infrared spectrum of the directly imaged planet HR 8799 e, finding an apparent degeneracy between solutions with significant cloud opacity, and those with less cloudy atmospheres but with much shallower temperature gradients. This reflects a theoretical prediction from \citet{Tremblin2015,Tremblin2016} that the red $J$--$H$ and $J$--$K$ colors of many L dwarfs may just as readily be explained by a chemo-convective instability that produces vertical temperature gradients shallower than would be expected in thermo-chemical equilibrium. \citet{Nowak2020} produce and compare retrievals from both the \texttt{ExoREM} and \texttt{petitRADTRANS} codes on an $R \sim 500$ $K$-band spectrum of $\beta$ Pictoris b, finding excellent agreement between the retrieved C/O ratios of the two codes. \citet{Wang2022}'s retrieval on $K$ band data of HR 7672 B represents the highest resolution spectrum used in an L dwarf retrieval to date, at $R \sim 35000$, and were able to precisely constrain the H$_2$O and CO abundances, finding a C/O ratio consistent with the primary. \citet{Lueber2022} present a systematic retrieval of brown dwarfs across the L and T spectral types at an average resolution $R \sim 100$ across the near-infrared, but do not find any consistency or trends in the retrieved cloud properties for the L dwarfs. However, when considering mid-infrared (here, 5--14 $\mu$m) spectra at similar resolution, \citet{Suarez2022} find silicate features emerge starting at a spectral type of approximately L2, continuing through the mid-Ls, with the variability of the brown dwarf correlating positively with the presence and strength of silicate absorption. In this work we will present our own retrieval efforts on a widely separated companion classified as an early L dwarf, and will use an additional retrieval on a previously-studied field L dwarf to compare the results of our code with those of a different retrieval code on an object in a similar spectral class.

We discuss the available data on the HD 106906 system and its companion in \S \ref{sec:data} and describe the components of our atmospheric forward model and retrieval code in \S \ref{sec:model}. We test the ability of our code to converge on consistent results by using synthetic data in \S \ref{sec:self-results}, benchmark the code by modeling a field L dwarf that has been previously retrieved on with a different code (\S \ref{sec:test-results}), and finally show our results for the L dwarf companion HD 106906 b in \S \ref{sec:results}. Finally, we discuss the interpretation and limitations in our retrieval in \S \ref{sec:discussion} and summarize our key findings in \S \ref{sec:conclusion}.

\section{The HD 106906 System}\label{sec:data}
\begin{deluxetable}{lcr}[htb]
\tabletypesize{\footnotesize}
\tablewidth{0pt}
\tablecaption{Fundamental properties for the HD 106906 system and companion HD 106906 b.}
\tablehead{\colhead{Name} & \colhead{Value} & \colhead{Reference}}
\startdata
Distance (pc) & $103.3\pm0.4$ & \citet{GaiaCollaboration2020} \\
Projected separation $\left(^{\prime\prime}\right)$ & $7.11\pm0.03$ & \citet{Bailey2014} \\
Projected separation (AU) & $734\pm4$ & Calculated from above. \\
Age (Myr) & $15\pm3$\tablenotemark{a} & \citet{Pecaut2012} \\
$M_\star/M_\odot$ (binary, total) & $2.58\pm0.04$ & \citet{Lagrange2016} \\
$M_\mathrm{comp}/M_\mathrm{J}$ & $11\pm2$ & \citet{Bailey2014} \\
$\log_{10}\!\left(L/L_\odot\right)$ & $-3.65\pm0.08$ & \citet{Daemgen2017} \\
$T_\mathrm{eff}$ (K) & $1820\pm240$ & \citet{Daemgen2017} \\
Spectral Type & L$1.5\pm1.0$ & \citet{Daemgen2017} \\
\enddata
\tablenotetext{a}{The age estimate for the Lower Centaurus Crux subgroup of the Scorpius-Centaurus OB association, of which the HD 106906 system is a member.}
\label{table:HD106906_parameters}
\end{deluxetable}

\begin{deluxetable}{lcr}[htb]
\tabletypesize{\footnotesize}
\tablewidth{0pt}
\tablecaption{Photometric and spectral properties for the companion HD 106906 b.}
\tablehead{\colhead{Name} & \colhead{Value} & \colhead{Reference}}
\startdata
\cutinhead{\textbf{$J$} (1.10--1.35 $\mu$m)}
Flux Density (10${^{-13}}$ erg s$^{-1}$ cm$^{-2}$ $\mu$m$^{-1}$) & $2.86_{-0.69}^{+0.91}$ & \multirow{2}{*}{\citet{Bailey2014}} \\
Magnitude (2MASS)     & $17.6\pm0.3$      &  \\
Magnitude (STMAG equivalent)     & $20.3\pm0.3$      & Calculated.\tablenotemark{b} \\
Resolution    & $\approx 2000$    & \multirow{2}{*}{\citet{Daemgen2017}} \\
S/N per pixel & $\approx 20$      & \\
\cutinhead{\textbf{$H$} (1.45--1.81 $\mu$m)}
Flux Density (10${^{-13}}$ erg s$^{-1}$ cm$^{-2}$ $\mu$m$^{-1}$) & $3.68_{-0.89}^{+1.17}$ & \multirow{2}{*}{Estimated.\tablenotemark{a}} \\
Magnitude (2MASS)     & $16.2\pm0.3$      &  \\
Magnitude (STMAG equivalent)     & $20.0\pm0.3$      & Calculated.\tablenotemark{b} \\
Resolution    & $\approx 3000$    & \multirow{2}{*}{\citet{Daemgen2017}} \\
S/N per pixel & $\approx 20$--50  & \\
\cutinhead{\textbf{$K\mathrm{s}$} (1.94--2.46 $\mu$m)}
Flux Density (10${^{-13}}$ erg s$^{-1}$ cm$^{-2}$ $\mu$m$^{-1}$) & $2.81_{-0.15}^{+0.16}$ & \multirow{2}{*}{\citet{Bailey2014}} \\
Magnitude (2MASS)     & $15.46\pm0.06$    & \\
Magnitude (STMAG equivalent)     & $20.28\pm0.06$    & Calculated.\tablenotemark{b} \\
Resolution    & $\approx 4000$    & \multirow{2}{*}{\citet{Daemgen2017}} \\
S/N per pixel & $\approx 20$--40  & \\
\cutinhead{{HST/WFC3/F127M} (centered at 1.274 $\mu$m)}
Magnitude (STMAG) & $19.41\pm0.01$ & \multirow{2}{*}{\citet{Zhou2020}} \\
Flux Density (10${^{-13}}$ erg s$^{-1}$ cm$^{-2}$ $\mu$m$^{-1}$) & $6.28\pm0.08$ & \\
\cutinhead{{HST/WFC3/F139M} (centered at 1.384 $\mu$m)}
Magnitude (STMAG) & $19.97\pm0.01$ & \multirow{2}{*}{\citet{Zhou2020}} \\
Flux Density (10${^{-13}}$ erg s$^{-1}$ cm$^{-2}$ $\mu$m$^{-1}$) & $3.74\pm0.05$ & \\
\cutinhead{{HST/WFC3/F153M} (centered at 1.532 $\mu$m)}
Magnitude (STMAG) & $19.79\pm0.01$ & \multirow{2}{*}{\citet{Zhou2020}} \\
Flux Density (10${^{-13}}$ erg s$^{-1}$ cm$^{-2}$ $\mu$m$^{-1}$) & $4.39\pm0.04$ & \\
\enddata
\tablenotetext{a}{Estimated from $J-H$ and $H-K_\mathrm{s}$ colors for low-gravity L dwarfs; see Table 3 in \citet{Faherty2013}.}
\tablenotetext{b}{The zero points of the 2MASS and STMAG systems differ. To properly compare the mean flux density of a spectrum across an HST bandpass where only a 2MASS magnitude is available, one should add $\approx 2.66$ to a 2MASS $J$ to get its equivalent STMAG; for $H$ and $K_\mathrm{s}$, add $\approx 3.78$ and 4.82 magnitudes, respectively.}
\label{table:HD106906b_parameters}
\end{deluxetable}

\subsection{System Properties}\label{sec:data:system}
The HD 106906 system consists of a pair of nearly identical-mass young F-type stars (combined mass $2.6M_\odot$) orbiting each other at $0.36\pm0.002$ AU \citep{Lagrange2016,Rodet2017,Rosa2019}. Its membership in the Lower Centaurus Crux (LCC) association \citep{Gagne2018} places the system's age at $15\pm3$ Myr \citep{Pecaut2012,Pecaut2016}. A companion, HD 106906 b, has an estimated mass of $11\pm2 M_\mathrm{J}$ from fits to evolutionary models and sits at a projected angular separation of $7^{\prime\prime}.11\pm0^{\prime\prime}.03$ \citep{Bailey2014}. At a distance of $103.3\pm0.4$ pc \citep{GaiaCollaboration2020}, this places HD 106906 b at a projected physical separation of $734\pm4$ AU \citep{Zhou2020}. 

Given its mass and orbit, HD 106906 b straddles the line between planets and brown dwarfs. Its exact formation pathway remains uncertain, as its mass ratio relative to --- and remarkably wide separation from --- its binary hosts pose challenges for all possible scenarios. To date, no studies of the HD 106906 system have provided substantial evidence for one formation pathway over another for the companion. On the planet formation side, there are several efforts to understand the dynamical history of the HD 106906 system, given the misalignment of the companion with the observed debris disk \citep{Bailey2014,Kalas2015,Lagrange2016,Bryan2021}. A core accretion pathway would require HD 106906 b to have formed interior to $\lesssim$100 AU, which then requires a mechanism to evolve its orbit to the current projected separation in excess of 700 AU. \citet{Wu2016} highlight the possibility that HD 106906 b could have been a planet scattered outward by its binary host, though the binary-planet scattering time scale is thought to be longer than the age of the system \citep{Jilkova2015}. This hypothesis has been tested with efforts to constrain its orbital motion \citep{Rodet2019,Nguyen2020}. \citet{Nguyen2020} posit that HD 106906 b's orbit could have been excited in both orbital eccentricity and inclination from an unstable resonance with the binary.\footnote{This mechanism is of great interest in understanding the formation and evolution of the purported ``Planet Nine'' in our own Solar System, and may serve as a general mechanism for explaining the observation of planetary-mass companions at orbital separations $\gtrsim 100$ AU.} However, this explanation is unlikely for HD 106906 in particular given the low density ($<0.11$ stars per cubic parsec) of the LCC, which makes it unlikely that the companion's current position is the result of fly-bys scattering an initially closer-in orbit. The most recent study of the dynamical origin of this system is found as of the writing of this article is \citet{Moore2023}, who provide an argument via numerical simulations that HD 106906 b could have been captured into the system as a planetary-mass free-floating object. They estimate the probability of this scenario occurring within the last 5 Myr is $\sim 10^{-6}$, which, while still low, is an order of magnitude more likely than in previous estimates.

\citet{Swastik2021} demonstrate that the occurrence rate of companions at $\sim$10--1000 AU shows a negative correlation with host metallicity -- as opposed to the positive correlation seen in close-in gas giants -- for masses greater than about 4 Jupiter masses. This suggests that the formation histories of both the most massive planets and brown dwarfs may be dominated by gravitational instability, as the theory of formation by instabilities in the disk predicts a negative correlation with host metallicity \citep[see e.g.][]{Helled2009}. \citet{Bryan2021} find that the spin axis of HD 106906 b, its orbital plane, and the plane of HD 106906's circumstellar disk are all mutually misaligned. They conclude that formation via gravitational instability is a plausible mechanism, as it is most consistent with misalignment across all 3 vectors. This scenario points to a C/O ratio consistent with the hosts, as this could occur either with gravito-turbulent instability or fragmentation of a self-gravitating turbulent cloud.

\subsection{Data}\label{sec:data:data}
Photometry for the hosts and companion span the optical (F606W from \citealp{Kalas2015} and $z'$ from \citealp{Wu2016}) through the thermal infrared \citep[$L'$, see][]{Bailey2014}. Two sources of photometry exist within the \JHK{} wavelength range; see Table \ref{table:HD106906b_parameters}. The first is from \citet{Bailey2014}, who published $J$ and $K_\mathrm{s}$ magnitudes from the Magellan Adaptive Optics (MagAO) Clio2 instrument. The second is from \citet{Zhou2020}, who observed the HD 106906 system in the F127M, F139M, and F153M bands of the Wide Field Camera 3 (WFC3) of the Hubble Space Telescope (HST). The F127M bandpass overlaps with the $J$ bandpass, as does F153M's bandpass with that of $H$, thus providing an independent (though not precisely congruent) comparison with our estimated $H$ magnitude. F139M's bandpass falls almost entirely within the gap between the $J$ and $H$ band data.

The highest resolution spectrum of HD 106906 b comes from \citet{Daemgen2017}, who present data obtained with the SINFONI integral field spectrograph on the Very Large Telescope (VLT). The data consist of 3 spectra in $J$, $H$, and $K_\mathrm{s}$, discontiguous with each other, with resolutions $\approx 2000$--4000 , at a S/N ratio $\sim 20$--50 (see Table \ref{table:HD106906b_parameters}). They derive an effective temperature of $1820\pm240$ K and a spectral type of L$1.5\pm1.0$ based on comparisons with classifications in \citet{Allers2013} and \citet{Bonnefoy2014}. \citet{Daemgen2017} also classify the gravity-sensitive features as most consistent with a very low gravity \citep[consistent with the ``$\gamma$'' class, as defined and used in e.g.][]{Kirkpatrick2005,Cruz2009,Allers2013,Faherty2016}. 

There are a few factors that affect the uncertainties in the reduced spectra. Firstly, because the observations in \citet{Daemgen2017} lack a reliable $H$ band magnitude, we must make an estimate for the $H$ magnitude. We choose to calculate $J-H$ and $H-K$ colors from a selection of low-gravity L dwarfs, published in Table 3 of \citet{Faherty2013}. From these we take a weighted average with HD 106906 b's known $J$ and $K_\mathrm{s}$ magnitudes \citep{Bailey2014} to obtain an estimate $H = 16.2\pm0.2$. Secondly, \citet{Daemgen2017} identify regions, mostly at the edges of each spectroscopic band, that suffer large overall telluric absorption, as well as isolated wavelength ranges within each band (though concentrated in the H band) that may suffer from systematic uncertainties from the removal of telluric hydrogen in the data reduction. The reduced \JHK{} spectrum is normalized to the flux density at a specific reference wavelength in each band.

A striking disagreement arises when comparing the flux densities inferred from the $J$ versus that of the F127M photometry: the flux density derived from HST photometry is roughly twice as bright. To calculate this we take the portion of our spectrum within the F127M filter, and calculate how bright this object would be given the $J$ magnitude, since the vast majority of the F127M band lies within the $J$ band. From this calculation (done using the \texttt{pysynphot} package, see \citealp{pysynphot}) we expect to see a F127M flux density of $\approx 3.15\times10^{-13}$ erg cm$^{-2}$ s$^{-1}$ $\mu$m$^{-1}$, versus the $\approx 6.28\times10^{-13}$ as actually was observed with HST and reported in \citet{Zhou2020}. L dwarfs are known to be variable from photometric monitoring, with the notable case of VHS 1256 b with a $\sim 20$\% variability across 1.1--1.7 $\mu$m with a period of $\approx 21$--24 hours \citep{Bowler2020}, and variability at the $\sim 6$\% level when extending to 5 $\mu$m \citep{Zhou2020a}. This prevailing hypothesis for this variability is rotation \citep[see also e.g.][]{Zhou2016}, with non-uniform cloud cover imparting brightness variations as changes in visible cloud opacity. However, \citet{Zhou2020} found that HD 106906 b was only variable at the $\sim 1$\% in the HST WFC3 F127M band, which is far smaller than would be needed to explain the discrepancy.

One may prefer to adopt the HST photometry as the point of reference for the $J$ and $H$ bands, as space-based photometry provides a greater instrumental precision and does not suffer from systematics from telluric subtraction. However, there are no direct HST comparisons for the $K_\mathrm{s}$ band, and so one will still need to mix the sources of photometry to calibrate the entire spectrum. We do not resolve the disagreement between these two sources of photometry --- instead, we run our retrieval of HD 106906 b with the original $J$ and $K_\mathrm{s}$ photometric normalization. We use calibration parameters in flux normalization to attempt to capture uncertainties in this flux normalization. These uncertainties stem from the error bars on the $J$ and $K_\mathrm{s}$ magnitudes used to normalize the spectra in \citet{Daemgen2017}.

We do not have a direct constraint on the C/O ratio of the hosts; such a constraint is needed to compare with the C/O ratio we retrieve to test the hypothesis that the HD 106906 b formed as a substellar companion. One approach is to combine a C/O to [Fe/H] relation for planet hosts with metallicity measurements for stars in the galaxy. In the case of HD 106906, one can use metallicities from members of the Upper Centaurus-Lupus (UCL) and Lower Centaurus-Crux (LCC) associations \citep[see Table 1 in][]{Bubar2011}, which yields a mean $\left[\mathrm{Fe}/\mathrm{H}\right] = -0.12\pm0.09$. The discussion of \citet{Nissen2013} provides a C/O to [Fe/H] relation for planet hosts:
\begin{equation}\label{eq:ctoo_feonh}
    \mathrm{C}/\mathrm{O} = 0.58 + 0.48\left[\mathrm{Fe}/\mathrm{H}\right],
\end{equation}
with an RMS dispersion $\sigma\!\left(\mathrm{C}/\mathrm{O}\right) = 0.06$. Using this relation with the \citet{Bubar2011} mean metallicity, we get an estimate for the typical C/O ratio for a member of the Sco-Cen association:
\begin{equation}\label{eq:ctoo_Sco-Cen}
    \mathrm{C}/\mathrm{O}_\mathrm{Sco-Cen} = 0.52 \pm 0.11
\end{equation}
which is consistent with both the solar C/O found in \citet{Nissen2013} (C/O$_\odot = 0.58$), as well as the range of $0.55\pm0.10$ given in \citet{Asplund2009}.

\section{Atmospheric Modeling and Spectral Retrieval}\label{sec:model}
To model the atmosphere of HD 106906 b from its emission spectrum we employ the \href{https://github.com/alexrhowe/APOLLO}{\APOLLO{}} code \citep{Howe2017,Howe2022}, a model framework for generating spectra of planets both in transit and emission.\footnote{https://github.com/alexrhowe/APOLLO} The core of the forward model is modeling the combination of thermal emission, absorption from atomic and molecular species, and extinction (scattering and absorption) from clouds. \APOLLO{} uses a hemispherical approximation to the \citet{Toon1989} 2-stream scattering routine, which is used primarily for cloud scattering. To calculate the emission spectrum, the outgoing radiation is averaged over 8 angle divisions in the outgoing hemisphere. The code is designed to be modular in parametrizations of molecular abundances, temperature-pressure, observing modes, and noise models, with particular focus on observing configurations for \JWST{}. \APOLLO{} is equipped with a likelihood sampling routine that serves as the retrieval component of our model (discussed in Section \S \ref{sec:model:sampling}). The parameters used in our retrievals, including the bounds on their priors for the parameter estimation routine, are listed in Table \ref{table:APOLLO_parameters}.

\begin{deluxetable}{lr}[htb]
\tabletypesize{\footnotesize}
\tablewidth{0pt}
\tablecaption{Free parameters and the range of priors for the nested sampling algorithm used in our models with the \APOLLO{} code. All priors are uniform within the listed bounds. The calibration factors are also only used for retrieval on HD 106906 b. CO$_2$ is not included as an absorber in the retrieval of the spectrum of 2M2224.}
\tablehead{\colhead{Name} & \colhead{Range of Prior}}
\startdata
\cutinhead{Fundamental}
$R/R_\mathrm{J}$ & 0.04 -- 4 \\
log g = $\log_{10}\!\left[g/\!\left(\mathrm{cm~s}^{-2}\right)\right]$ & 2.5 -- 7.5 \\
\cutinhead{Gases ($\log_{10}$ number abundance relative to total; see \S \ref{sec:model:molecular})}
\vspace{-0.2cm} H$_2$O, CO, CO$_2$, H$_2$S & \\ \vspace{-0.2cm}
& $-12$ to $-1$ \\
Na+K, CrH, FeH, TiO, VO & \\
\cutinhead{Temperature-Pressure (see \S \ref{sec:model:T-P})}
Temperature at $10^{0.5}$ bars ($T_{0.5}$, K) & 75 -- 4000 \\
\vspace{-0.2cm} $T_{-4}$, $T_{-3}$, $T_{-2}$, $T_{-1}$, $T_{-0}$, & \\ \vspace{-0.2 cm}
& 75 -- 4000\tablenotemark{a} \\
$T_{1}$, $T_{1.5}$, $T_{2}$, $T_{2.5}$ (K) & \\
\cutinhead{Clouds (see \S \ref{sec:model:clouds})}
Power-law exponent $\left(\alpha\right)$ & $-10$ to 10 \\
$\log_{10}\!\left( P_\mathrm{top}/\mathrm{bar} \right)$ & $-4$ to 2.5 \\
$\log_{10}\!\left( \Delta P_\mathrm{cloud}/\mathrm{bar} \right)$ & 0 -- 6.5\tablenotemark{b} \\
Reference optical depth ($\log_{10}\tau\!\left(1\,\mu\mathrm{m}\right)$) & $-3$ to 2 \\
Single-scattering albedo $\left(\omega_0\right)$ & 0 -- 1 \\
Cloud filling fraction ($f$) & 0 -- 1 \\
\cutinhead{Calibration (see \S \ref{sec:model:sampling})}
Flux normalization in $J$ ($\Delta J$) & 0.5 -- 1.5 \\
Flux normalization in $H$ ($\Delta H$) & 0.5 -- 1.5 \\
\enddata
\tablenotetext{a}{The temperature profile is also constrained to be monotonic; see the discussion on the dependent priors in \S \ref{sec:model:T-P}.}
\tablenotetext{b}{An additional constraint is imposed such that the cloud layer does not extend beyond the base of the model.}
\label{table:APOLLO_parameters}
\end{deluxetable}

\subsection{Molecular Species and Opacities}\label{sec:model:molecular}
The molecular radiative transfer scheme in \APOLLO{} relies on sampling of cross-section tables for a variety of applicable species; these are pre-computed from a grid of line-by-line opacities and are derived from the sources in Table 1 of \citet{Freedman2014}, with the exception of the alkalis. The opacities for the alkali lines are drawn from the \citet{Lupu2022} catalog, which derives its Na and K profiles from a series of works analyzing the interaction between atomic lines and molecular hydrogen \citep{Allard2007,Allard2012a,Allard2016,Allard2019}. We employ two levels of down-sampling to create the cross-sections we use in our forward models, with resolutions of 10000 and 50000. For retrievals on real data, we choose the minimum opacity resolution that ensures the ratio between the mean opacity and data resolution is $\gtrsim 100$, following a community recommendation to avoid introducing artificial errors from binning effects.\footnote{See, for example, the \href{https://natashabatalha.github.io/picaso/notebooks/10_CreatingOpacityDb.html}{discussion on opacity resampling} in the \texttt{PICASO} documentation.} We freely retrieve fractional abundances for H$_2$O, CO, CO$_2$, H$_2$S, CrH, FeH, TiO, and VO. We assume a solar H/He ratio, and H$_2$ and He opacities include collisionally-induced absorption (CIA). Atomic Na and K are included together as a single free parameter, where the ratio of their abundances is fixed to that of solar metallicity \citep[see e.g.][]{Line2015}. For the molecular abundances we assume a constant mixing ratio, and initialize at values corresponding to the chemical equilibrium abundances at the pressure layer closest to the literature effective temperature \citep[here taken to be 1820 K from][]{Daemgen2017}. The equilibrium abundances were calculated through a routine in the \href{https://natashabatalha.github.io/picaso_dev}{\texttt{PICASO}} atmospheric radiative transfer code \citep{Batalha2019}.

We visualize the contributions of the gas to the emission spectrum by calculating a contribution function per atmospheric layer, which is given by
\begin{equation}\label{eq:contribution_function}
    C_\mathrm{sp}\!\left(P, \lambda\right) \equiv B_{\lambda}\!\left(T\!\left(P\right)\right) \frac{\int_{P}^{P+\Delta P}\,d\tau_\mathrm{sp}}{\exp\!\left(\int_{0}^{P+\Delta P}\,d\tau_\mathrm{tot}\right)}
\end{equation}
where the atmospheric layer spans pressures $P$ to $P+\Delta P$, $T\!\left(P\right)$ is the temperature in the layer, and $B_{\lambda}$ is the Planck function at that temperature. $\tau_\mathrm{sp}$ and $\tau_\mathrm{tot}$ represent the optical depths due to a given gas species and from the entire contents of the layer, respectively. The contribution function is expressed as fractions of the total across an entire vertical column in the atmosphere. The function, when summed across all gas and cloud species, is proportional to the pressure derivative of the ``transmittance'' ($\exp\!\left(-\tau\right)$) times the Planck function at the given pressure and temperature; see for example \citet{Line2014}, \S 3.

\subsection{Temperature-Pressure Profile}\label{sec:model:T-P}
Our T-P profile is adapted from the parametrization proposed in \citet{Piette2020}, Section 4.2 (see Figure 8 in their paper), with additional temperature nodes added to the extremes of the profile. The parametrization is designed to be flexible enough to accommodate a wide range of possible vertical thermal structures, including approximation a radiative-convective equilibrium, while also filtering out excessive unphysical behavior, such as the ``ringing'' that was described in \citet{Line2015,Line2017}. The parameters are the temperatures at 10 pressure levels, representing nodes between which we interpolate the profile. The temperature nodes are spaced in orders of magnitude from the top of the model atmosphere ($10^{-4}$ bar) down to a pressure of 1 bar, beyond which we use half-orders until we reach the deepest pressure of the model at $10^{2.5}$ bars. Here we label the temperatures of each node by subscripts denoting the base-10 logarithm of their corresponding pressure. We follow the recommendation of \citet{Piette2020} to use a monotonic spline interpolation with a Gaussian smoothing kernel of width 0.3 dex in log-pressure, as the mechanism by which one can filter out the aforementioned ringing. In the original setup, the temperature at a pressure of $10^{0.5}$ bars ($T_{0.5}$) is taken as a reference temperature at a fiducial pressure approximating the depth of the photosphere for a typical self-luminous brown dwarf. In this setup, the remaining parameters then define the \emph{differences} in temperature between each successive node. In contrast, we choose to define all our parameters as the temperatures themselves, but use an iterative process for proposing temperatures by determining the bounds on the uniform priors for each temperature:
\begin{itemize}
    \item The bounds of the prior for the photospheric node ($T_{0.5}$) are set by the bounds of the temperatures of the opacity tables (75--4000 K).
    \item Then, the shallowest ($T_{-4}$) and deepest ($T_{2.5}$) temperature prior bounds are each bounded by the proposal for $T_{0.5}$ and by the minimum and maximum opacity temperatures, respectively.
    \item This continues with the nodes closest to the middle of the existing nodes being bounded by those already chosen nodes, sub-dividing until the whole profile is bounded and all temperatures proposed.
\end{itemize}
This ensures the profile is monotonic in temperature.

\subsection{Cloud Models}\label{sec:model:clouds}
Our cloud model is modeled after the ``slab'' approaches used in \citet{Burningham2017} and \citet{Gonzales2020}. The model cloud occupies a fixed region in pressure space, with a minimum pressure where cloud absorption begins (the cloud ``top''), and some depth in pressure. The vertical opacity profile is restricted to follow $\partial\tau/\partial P \propto P$. The free parameters include the pressure of the cloud top $P_\mathrm{top}$, the depth of the cloud in log-pressure space $\log_{10}\!\left( \Delta P_\mathrm{cloud}\right) \equiv \log_{10}\!\left(P_\mathrm{base}/P_\mathrm{top}\right)$, and a wavelength-dependent opacity and single-scattering albedo instead of particle-specific parameters. The wavelength dependence is modeled as a power law with exponent $\alpha$. The opacity at a given pressure depth and wavelength is therefore given as
\begin{equation}\label{}
    \tau\!\left(P, \lambda\right) = \tau_0 \left(\frac{\lambda}{\mu\mathrm{m}}\right)^\alpha \left(\frac{P^2 - P^2_\mathrm{top}}{P^2_\mathrm{base} - P^2_\mathrm{top}}\right)
\end{equation}
where $\tau_0$ is the maximum optical depth (at the base of the cloud at a pressure $P_\mathrm{base}$) at a wavelength of 1 $\mu$m. This is an empirical approximation to scattering by cloud particles whose sizes are smaller than the wavelengths of observation. Previous efforts at retrievals in this wavelength range indicate that a model that takes into account specific condensates for its opacity calculations is not preferred over the simpler approach used in this work.

The final free parameter for our cloud model is the single-scattering albedo $\omega_0$. This is the ratio of photons that are scattered versus those extincted overall (either absorbed or scattered). By choosing to model the albedo with a single free parameter, we assume it is constant across all wavelengths and pressures. To be precise, $\omega_0$ in this work refers to the single-scattering albedo of the clouds alone; in \APOLLO{}'s implementation of the \citet{Toon1989} radiative transfer model, their $\omega_0$ refers more broadly to the scattering-to-extinction ratio of all absorbers and scatterers in the atmosphere. For our purposes this means including the gas as well, for which we model scattering as Rayleigh scattering since the sizes of each molecule are much smaller than the observed wavelengths.

\subsection{Parameter Estimation Methods}\label{sec:model:sampling}
We sample likelihoods in parameter space with a nested sampling algorithm, using the \texttt{dynesty} Python package \citep{Speagle2020}. We choose to set uniform priors on all parameters, the ranges of which are listed in Table \ref{table:APOLLO_parameters}. Each model was initialized with 1000 live points in the ``\href{https://dynesty.readthedocs.io/en/stable/api.html#dynesty.dynesty.DynamicNestedSampler}{\texttt{rwalk}}'' sampling method. Models were run with the built-in default stopping criterion for assessing convergence, which depends on the amount of evidence accounted for in the cumulative samples.\footnote{See the \href{https://dynesty.readthedocs.io/en/stable/index.html}{\texttt{dynesty} documentation} for more information on how stopping criteria are applied.} The total number of effective iterations in each run varies based on when the stopping criteria are reached, with test retrievals on simulated data (\S \ref{sec:self-results}) using $\sim 10^5$, and retrievals on real data (\S \ref{sec:test-results} and \ref{sec:results}) requiring 2--3 times as many.

Once the runs are complete, we then derive the mass, effective temperature, metallicity, and C/O ratio. The mass is calculated directly from the radius and surface gravity. The effective temperature is calculated from an approximation to the bolometric luminosity, using a low-resolution ($R \approx 200$) spectrum that covers 0.6--30 $\mu$m\footnote{Note that for forward models, especially those with negative cloud opacity power-law exponents, two spectra can have considerably different effective temperatures while only displaying modest differences in the spectra in the near-infrared. This is discussed briefly in the section on our self-retrievals on cloud-free simulations (\S \ref{sec:self-results:results:cloud-free}).}. We report metallicity by comparing the mean molecular weight versus that expected for solar metallicity, rather than reporting a metallicity as an [Fe/H] value. For the mass fraction $Z$ of non-H/He elements, the metallicity is calculated as $\log_{10}\!\left(Z/Z_\odot\right)$, where we take $Z_\odot = 0.0196$. We choose this definition for metallicity because our values of metallicity are not tied to a specific atomic species, and the way in which we model the abundances --- uniform in pressure but freely variable --- means our model does not require the abundances to be in chemical equilibrium.

We use Bayes factors to compare the quality of fits to data between two models. The Bayes factor is simply the ratio of the marginal likelihoods (also known as evidences) of each model's retrieval. A higher Bayes factor confers stronger support for a model relative to another; a recommendation originally proposed in \citet{Jeffreys1998} is to interpret a ratio of 10--$10^{1.5}$ as ``strong'', $10^{1.5}$--$10^2$ as ``very strong'', and $>10^2$ as ``decisive'' confidence that the model with the higher evidence is preferred. Following this, \citet{Benneke2013} adapted the heuristics in Table 1 of \citet{Trotta2008} to translate the language of Bayes factor comparisons into a ``detection significance'', usually quoted in units of ``sigma'' $\sigma$. This is a convenient way to express analogous statistics in both the Bayesian and frequentist frameworks of model analysis; we report both in the following sections for our model comparisons.

\section{Retrieving on Simulated Data from Forward Models of Low-Gravity L Dwarfs}\label{sec:self-results}
This work represents the first application of \APOLLO{} to data from an L dwarf. To test the efficacy of the code, we generate a forward model that approximates the object, making simple assumptions about the atmospheric structure. In principle our retrieval code should be able to converge on a good fit (i.e.~reduced chi-square statistic $\chi_\nu^2 \sim 1$) to a dataset generated from its own forward model, and, based on the distributions of the retrieved parameters, should inform us to how well each parameter could be constrained from a near-infrared wavelength range and signal-to-noise similar to that of HD 106906 b.

We use \APOLLO{} to generate a forward model spectrum for a 1.5 $R_\mathrm{J}$, $\log g = 4.19$ object; the choice of radius is arbitrary but the surface gravity is taken from the best estimate of HD 106906 b's gravity from the observed luminosity and effective temperature based on the fits made in \citet{Daemgen2017}. For the thermal profile, we produce a parametrization that approximates a \texttt{SONORA} profile at approximately 1800 K if no cloud opacity is present. We use \texttt{PICASO}, specifically the Visscher chemical equilibrium code \citep{Marley2021}, to generate equilibrium abundances for the model pressures given the above parameters. This corresponds to a C/O ratio of 0.54 and a metallicity of 0.065. We generate data for two cases: one with clouds, parametrized as described in \S \ref{sec:model:clouds}, and one ``clear'' case without clouds. For the clouds, we use a layer that spans $\sim 10^{-0.5}$--$10^{1}$ bars, chosen to bound the estimated photospheric pressures, and has enough opacity to yield an effective temperature of $\approx 1360$ K. All models used to generate these data are identical in all non-cloud parameter values. Noise is modeled as independent, Gaussian (white) noise at S/N = 20, approximately the minimum S/N seen in the HD 106906 b spectrum. Comparisons of the forward model spectral fits, T-P profiles, and distributions of model parameters are shown in Figures \ref{fig:self-retrieval_on-clear_spectrum}--\ref{fig:self-retrieval_on-clear_cloud-corners} (on cloud-free data) and \ref{fig:self-retrieval_on-cloudy_spectrum}--\ref{fig:self-retrieval_on-cloudy_cloud-corners} (on cloudy data).

\begin{deluxetable}{lccccccccc}
\tabletypesize{\scriptsize}
\tablewidth{0pt}
\tablecaption{Median and MLE parameter values for the retrievals on simulated data, as described in \S \ref{sec:self-results}. The 2 models (cloud-free and cloudy) used to retrieve on the simulated data were the same as those used to generate the cloud-free and cloudy data. Both sets of parameter values are identical in all non-cloud parameters, and the values used in generating the simulated data are also identical except for the inclusion of clouds. We show results from each possible pairing of clear and cloudy model versus clear and cloudy data.}
\tablehead{ {} &
            {} &
           \multicolumn{2}{c}{Clear Model, Clear Data} &
           \multicolumn{2}{c}{Cloudy Model, Clear Data} &
           \multicolumn{2}{c}{Cloudy Model, Cloudy Data} &
           \multicolumn{2}{c}{Clear Model, Cloudy Data} \\
           \colhead{Name} &
           \colhead{True Value} &
           \colhead{Median} &
           \colhead{MLE} &
           \colhead{Median} &
           \colhead{MLE} &
           \colhead{Median} &
           \colhead{MLE} &
           \colhead{Median} &
           \colhead{MLE}}
\startdata
\cutinhead{Fit Quality}
$\chi_\nu^2$ & & & 1.00 & & 1.00 & & 1.01 & & 1.07 \\
\cutinhead{Fundamental}
$R/R_\mathrm{J}$     & 1.500 & $1.502\pm0.003$ & 1.504 & $1.515^{+0.021}_{-0.011}$ & 1.51 & $1.49\pm0.01$ & 1.49 & $1.315\pm0.003$ & 1.313 \\
$\log_{10}\!\left[g/\!\left(\mathrm{cm~s}^{-2}\right)\right]$ & 4.19 & $4.18\pm0.01$ & 4.18 & $4.17\pm0.01$ & 4.17 & $4.19\pm0.01$ & 4.20 & $4.39\pm0.01$ & 4.40 \\
$M/M_\mathrm{J}$     & 14.02 & $13.63^{+0.24}_{-0.20}$ & 13.64 & $13.63^{+0.33}_{-0.26}$ & 13.60 & $13.99^{+0.28}_{-0.29}$ & 14.17 & $17.24\pm0.30$ & 17.23 \\
$T_\mathrm{eff}$ (K) & \tablenotemark{a} & $1820^{+1}_{-2}$ & 1822 & $1274^{+499}_{-229}$ & 1813 & $1223^{+138}_{-158}$ & 1313 & $1872\pm1$ & 1871 \\
C/O                  & 0.54 & $0.545\pm0.003$ & 0.547 & $0.545^{+0.003}_{-0.005}$ & 0.547 & $0.535^{+0.004}_{-0.005}$ & 0.534 & $0.532\pm0.004$ & 0.531 \\
Metallicity          & 0.065 & $0.064^{+0.006}_{-0.007}$ & 0.065 & $0.063^{+0.008}_{-0.007}$ & $0.066$ & $0.055^{+0.009}_{-0.009}$ & 0.059 & $0.119^{+0.007}_{-0.008}$ & 0.117 \\
\cutinhead{Gases ($\log_{10}$ number abundance)}
H$_2$O   & $-3.35$ & $-3.358^{+0.006}_{-0.005}$ & $-3.359$ & $-3.357^{+0.006}_{-0.007}$ & $-3.359$ & $-3.354\pm0.007$ & $-3.348$ & $-3.288\pm0.006$ & $-3.287$ \\
CO       & $-3.28$ & $-3.279^{+0.007}_{-0.008}$ & $-3.278$ & $-3.280^{+0.010}_{-0.008}$ & $-3.276$ & $-3.293\pm0.011$ & $-3.288$ & $-3.231^{+0.008}_{-0.009}$ & $-3.234$ \\
CO$_2$   & $-7.00$ & $-8.85^{+1.86}_{-1.84}$ & $-8.24$ & $-8.93^{+1.36}_{-1.65}$ & $-8.65$ & $-8.49^{+1.39}_{-1.63}$ & $-8.50$ & $-8.87^{+1.62}_{-1.93}$ & $-8.88$ \\
H$_2$S   & $-4.60$ & $-4.56\pm0.03$ & $-4.56$ & $-4.57\pm0.02$ & $-4.55$ & $-4.60\pm0.03$ & $-4.64$ & $-4.52^{+0.04}_{-0.03}$ & $-4.52$ \\
Na+K     & $-5.42$ & $-5.423^{+0.005}_{-0.004}$ & $-5.428$ & $-5.423^{+0.011}_{-0.009}$ & $-5.421$ & $-5.438\pm0.010$ & $-5.432$ & $-5.369\pm0.006$ & $-5.371$ \\
CrH      & $-9.00$ & $-9.00^{+0.03}_{-0.02}$ & $-9.00$ & $-8.99^{+0.02}_{-0.03}$ & $-9.02$ & $-8.97\pm0.03$ & $-8.99$ & $-8.87\pm0.03$ & $-8.89$ \\
FeH      & $-9.00$ & $-9.01\pm0.02$ & $-9.01$ & $-9.02\pm0.02$ & $-9.01$ & $-9.01^{+0.02}_{-0.03}$ & $-8.99$ & $-8.96\pm0.03$ & $-8.94$ \\
TiO      & $-8.00$ & $-8.00\pm0.02$ & $-8.01$ & $-8.01^{+0.01}_{-0.02}$ & $-8.01$ & $-7.97\pm0.02$ & $-7.97$ & $-7.88\pm0.02$ & $-7.88$ \\
VO       & $-8.33$ & $-8.334^{+0.008}_{-0.009}$ & $-8.340$ & $-8.337\pm0.007$ & $-8.336$ & $-8.339^{+0.011}_{-0.010}$ & $-8.319$ & $-8.267\pm0.010$ & $-8.265$ \\
\cutinhead{Temperature-Pressure}
$T_{-4}$  (K)  &  723 & $622^{+205}_{-286}$ & 398 & $811^{+46}_{-148}$ & 782 & $606^{+84}_{-130}$ & 557 & $647^{+113}_{-268}$ & 609 \\
$T_{-3}$  (K)  &  826 & $848^{+41}_{-49}$ & 832 & $867^{+26}_{-36}$ & 854 & $779^{+35}_{-41}$ & 794 & $769^{+47}_{-56}$ & 775 \\
$T_{-2}$  (K)  &  964 & $962\pm4$ & 963 & $962^{+5}_{-6}$ & 961 & $948\pm6$ & 951 & $968^{+6}_{-7}$ & 969 \\
$T_{-1}$  (K)  & 1175 & $1175\pm2$ & 1175 & $1174^{+2}_{-3}$ & 1175 & $1176\pm2$ & 1176 & $1174\pm2$ & 1175 \\
$T_{0}$   (K)  & 1954 & $1958^{+2}_{-3}$ & 1956 & $1958^{+3}_{-2}$ & 1957 & $1948^{+3}_{-4}$ & 1950 & $1930\pm3$ & 1932 \\
$T_{0.5}$ (K)  & 2545 & $2546^{+4}_{-3}$ & 2546 & $2548\pm5$ & 2547 & $2548\pm8$ & 2560 & $2487\pm4$ & 2492 \\
$T_{1}$   (K)  & 3333 & $3243^{+36}_{-30}$ & 3271 & $3271^{+26}_{-32}$ & 3260 & $3095^{+51}_{-37}$ & 3294 & $2950^{+20}_{-27}$ & 2991 \\
$T_{1.5}$ (K)  & 4000 & $3377^{+178}_{-116}$ & 3548 & $3509^{+137}_{-157}$ & 3447 & $3095^{+51}_{-37}$ & 3294 & $2950^{+20}_{-27}$ & 2991 \\
$T_{2}$   (K)  & 4000 & $3568^{+198}_{-173}$ & 3692 & $3612^{+141}_{-189}$ & 3471 & $3388^{+180}_{-146}$ & 3488 & $3424^{+217}_{-224}$ & 3788 \\
$T_{2.5}$ (K)  & 4000 & $3702^{+203}_{-164}$ & 3790 & $3726^{+136}_{-150}$ & 3684 & $3515^{+193}_{-178}$ & 3667 & $3553^{+250}_{-240}$ & 3814 \\
\cutinhead{Clouds}
$\alpha$                                                         & -2.00 & & & $-0.94^{+1.78}_{-2.61}$ & $-0.38$ & $-2.21^{+0.41}_{-0.47}$ & $-1.66$ \\
$\log_{10}\!\left( P_\mathrm{top}/\mathrm{bar} \right)$          & -0.50 & & & $-1.03^{+2.05}_{-1.75}$ & $-0.59$ & $-1.02^{+0.49}_{-0.63}$ & $-0.67$ \\
$\log_{10}\!\left( \Delta P_\mathrm{cloud}/\mathrm{bar} \right)$ &  1.51 & & & $1.11^{+0.51}_{-0.58}$ & 1.00 & $1.43^{+0.61}_{-0.49}$ & 1.46 \\
$\log_{10}\!\left[\tau\!\left(\lambda_0\right)\right]$               &  0.48 & & & $-1.91^{+1.23}_{-0.69}$ & $-2.46$ & $-0.60^{+0.15}_{-0.11}$ & 0.13 \\
$\omega_0$                                                       &  0.66 & & & $0.07^{+0.05}_{-0.04}$ & 0.05 & $0.64\pm0.02$ & 0.63 \\
\enddata
\tablenotetext{a}{1821 K for the cloud-free data, and 1361 K for the data with power-law cloud opacity.}
\label{table:self-retrieval_best-fit_parameters}
\end{deluxetable}

\subsection{Retrievals on Cloud-free Data}\label{sec:self-results:results:cloud-free}
\begin{figure*}[htb]
\begin{center}
\textbf{Self-Retrieval on L Dwarf-like Cloud-Free Spectrum}\\
\includegraphics[width=17cm,trim={0 11.25cm 0 0},clip]{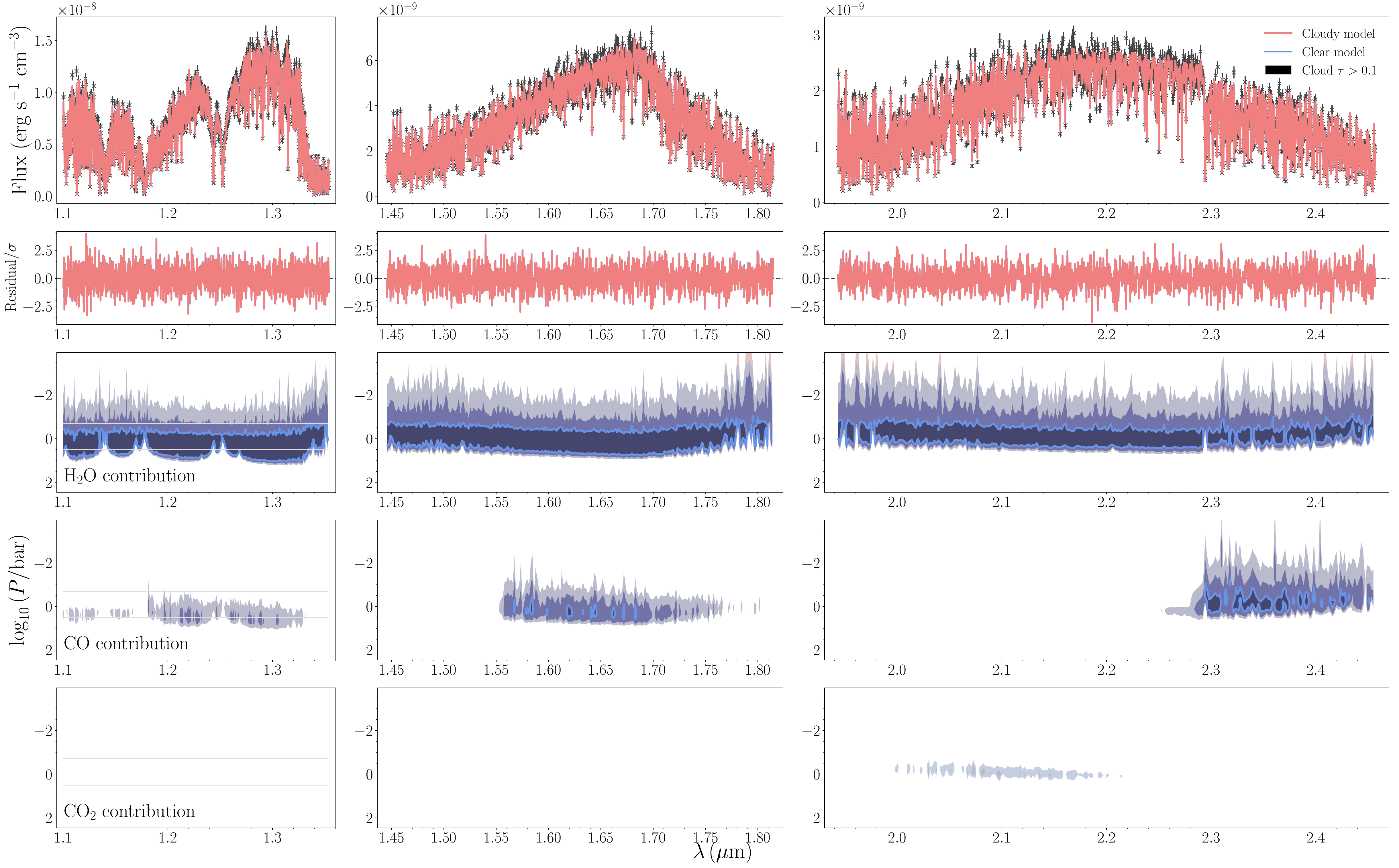}\\
$\lambda$ ($\mu$m)
\caption{Spectra and contribution functions of the retrieved forward model fits to data simulated from \textbf{cloud-free} forward models of an L dwarf. The forward model spectra (using the MLE parameter values) are in color, with the data in grey. The retrieved models are nearly identical and overlap nearly entirely. Immediately beneath the spectra are the contribution functions for the 2 principal carbon and oxygen bearing species; CO$_2$ is included in the model used to generate the simulated data, but its contributions to the emission are well below those of H$_2$O and CO. The deepest contours, outlined in solid colors, enclose the regions where the contribution function reaches $>1$\% of the total contribution within the atmospheric column at a given wavelength bin. Each successive contour denotes 2 orders of magnitude smaller fractional contribution (here, $10^{-4}$ and $10^{-6}$). The faint grey lines in the $J$-band (leftmost) sections contribution plots denote the location of the cloud layer as retrieved by the cloudy model on the cloud-free data. The faintness of the lines denotes the low optical depth of the cloud layer, in contrast with the darker cloud contours as seen in the retrieval on cloudy simulated data (Figure \ref{fig:self-retrieval_on-cloudy_spectrum}).}
\label{fig:self-retrieval_on-clear_spectrum}
\end{center}
\end{figure*}

\begin{figure*}[htb]
\begin{center}
\textbf{Self-Retrieval on L Dwarf-like Cloud-Free Spectrum}\\
\includegraphics[width=13cm]{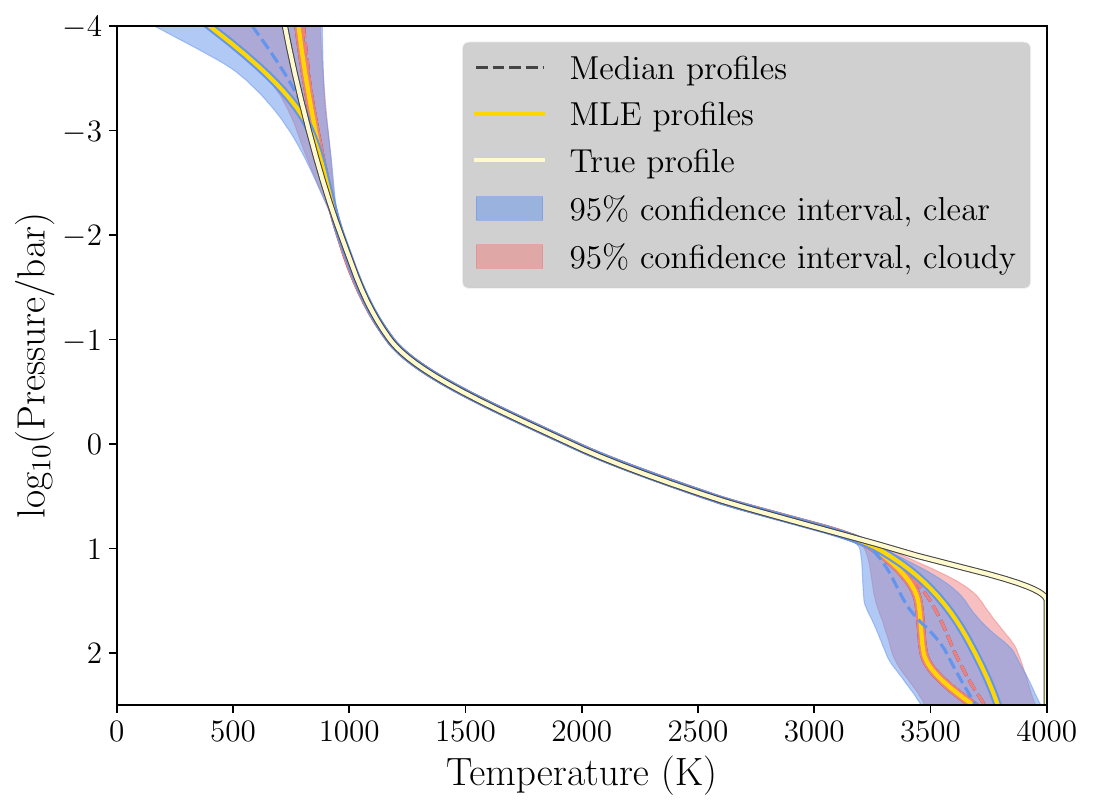}
\caption{The vertical temperature-pressure profiles of the retrieved forward model fits to data simulated from \textbf{cloud-free} forward models of an L dwarf. We show the MLE, median, and 95\% confidence interval of the retrieved T-P profiles with the true profile over-plotted.}
\label{fig:self-retrieval_on-clear_T-P}
\end{center}
\end{figure*}

\begin{figure*}[htb]
\begin{center}
\textbf{Self-Retrieval on L Dwarf-like Cloud-Free Spectrum}\\
\includegraphics[width=17cm]{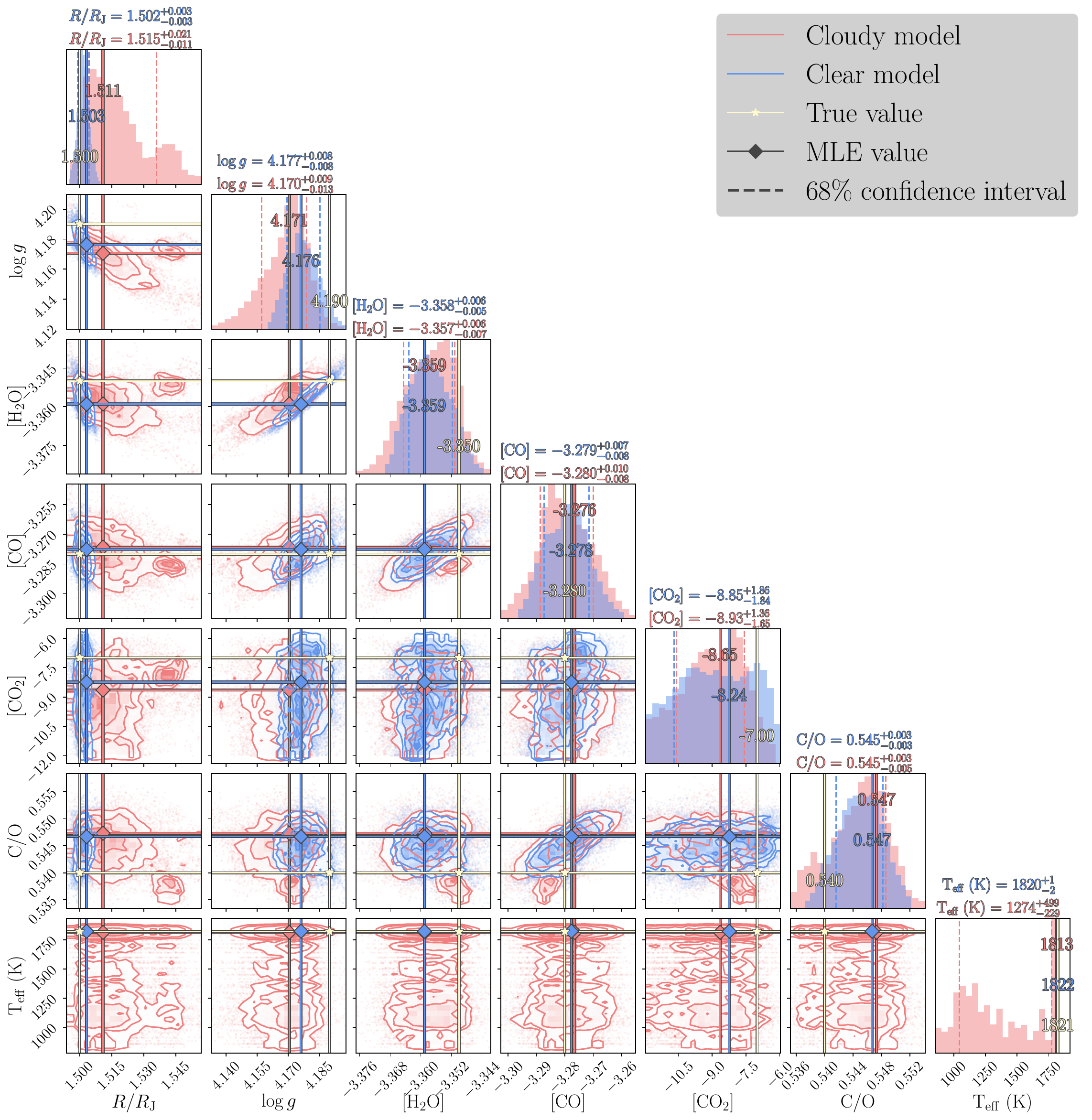}
\caption{A selection of parameters of the retrieved forward model fits to data simulated from \textbf{cloud-free} forward models of an L dwarf, shown as 1-D and 2-D histograms in a corner plot of the retrieved posterior distributions. The median value and 68\% confidence interval of each parameter are shown at the top of each column; the full list of median, interval range, and MLE values are shown in Table \ref{table:self-retrieval_best-fit_parameters}.}
\label{fig:self-retrieval_on-clear_custom-corners}
\end{center}
\end{figure*}

\begin{figure*}[htb]
\begin{center}
\textbf{Self-Retrieval on L Dwarf-like Cloud-Free Spectrum}\\
\includegraphics[width=17cm]{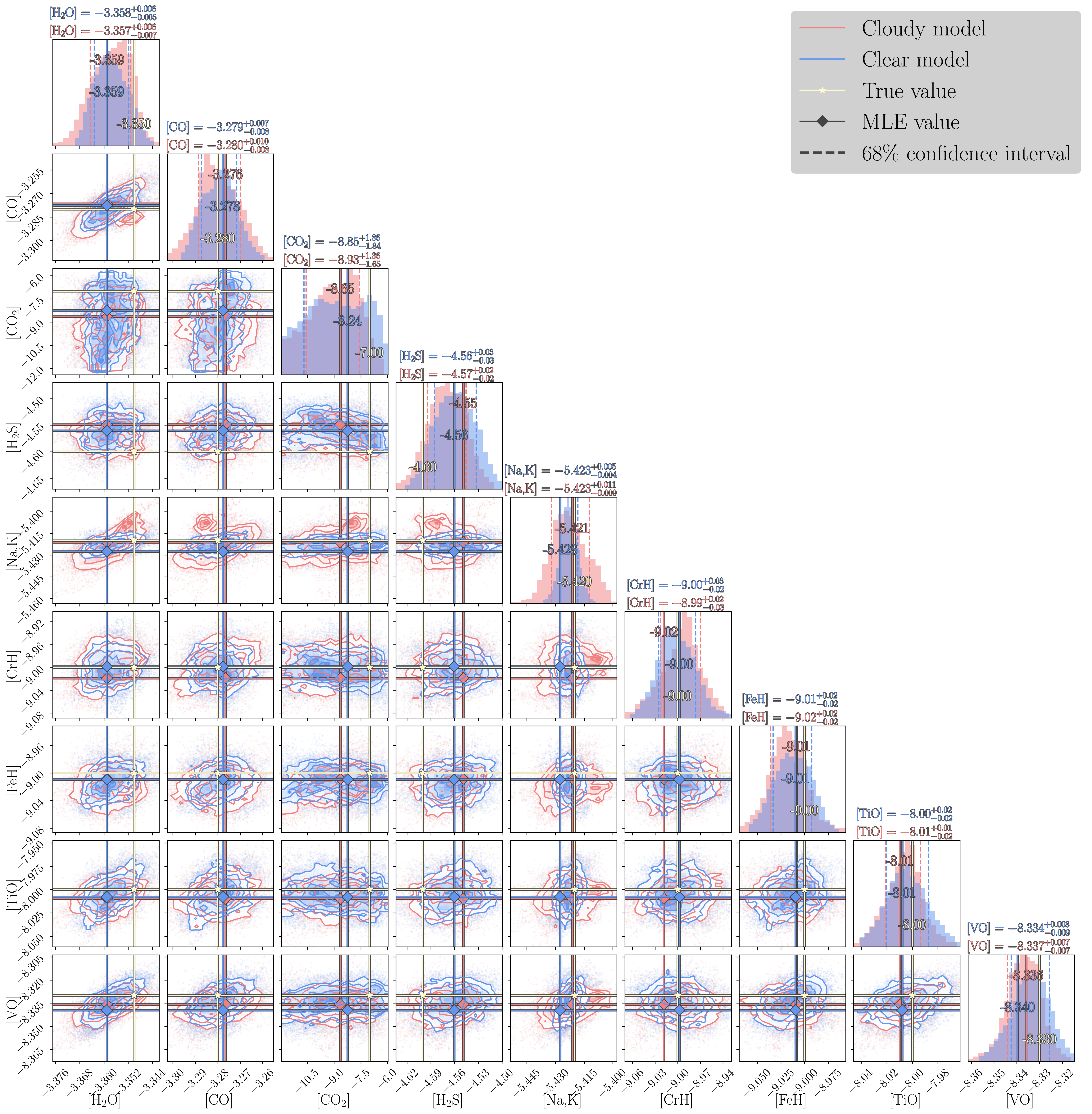}
\caption{Single-parameter (1-D) and parameter-versus-parameter (2-D) posterior distributions of gas parameters from the samples in the cloudy (red) and cloud-free (blue) forward model fit to data simulated from a \textbf{cloud-free} forward model of an L dwarf (see \S \ref{sec:self-results}). The cloud opacity is modeled as a power law in wavelength, as described in \S \ref{sec:model:clouds}. The median value and 68\% confidence interval of each parameter are shown at the top of each column; the full list of median, interval range, and MLE values are shown in Table \ref{table:self-retrieval_best-fit_parameters}.}
\label{fig:self-retrieval_on-clear_gas-corners}
\end{center}
\end{figure*}

\begin{figure*}[htb]
\begin{center}
\textbf{Self-Retrieval on L Dwarf-like Cloud-Free Spectrum}\\
\includegraphics[width=17cm]{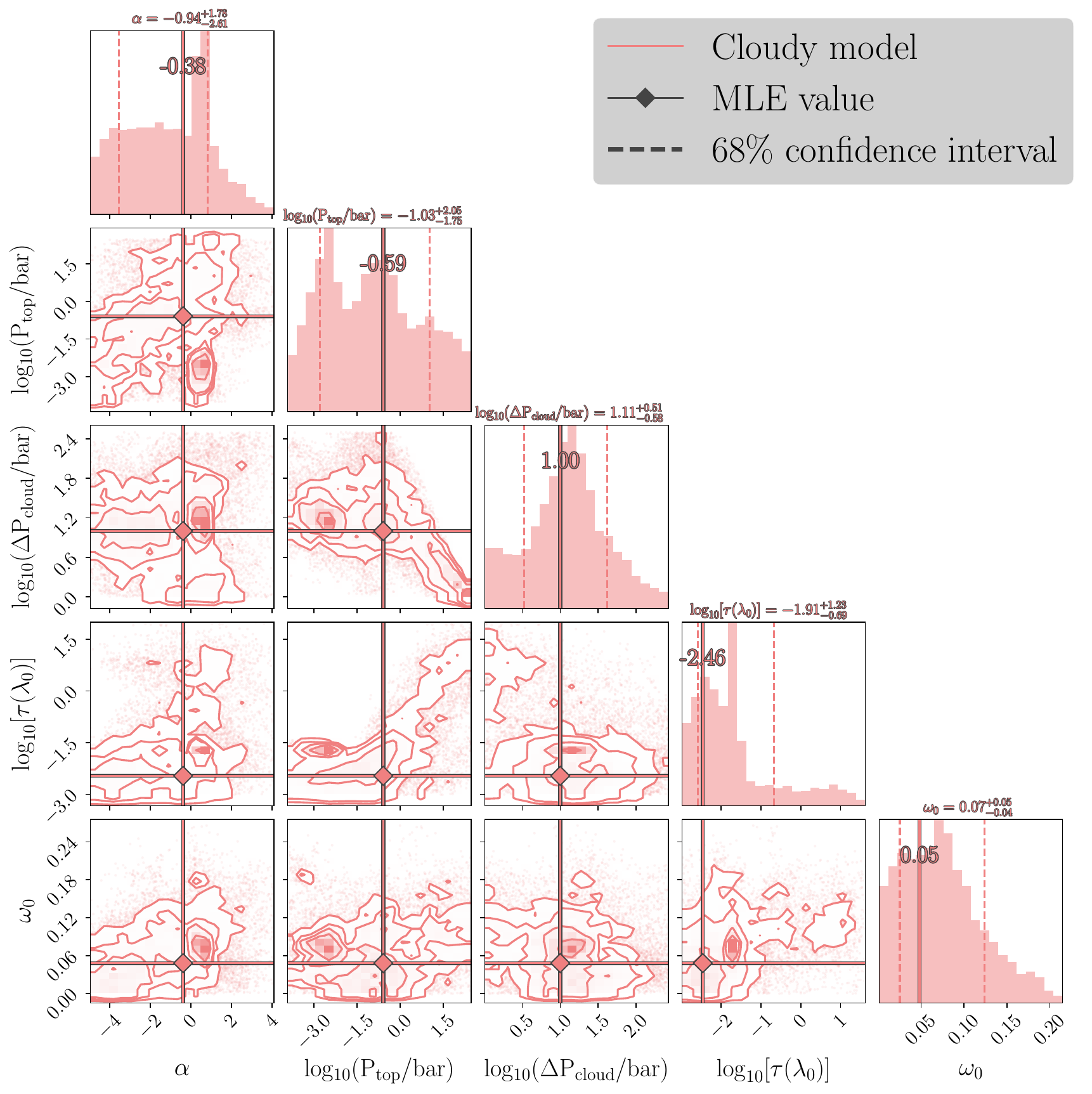}
\caption{Single-parameter (1-D) and parameter-versus-parameter (2-D) posterior distributions of cloud parameters from the samples in the cloudy forward model fit to data simulated from a \textbf{cloud-free} forward model of an L dwarf (see \S \ref{sec:self-results}). The cloud opacity is modeled as a power law in wavelength, as described in \S \ref{sec:model:clouds}. The median value and 68\% confidence interval of each parameter are shown at the top of each column; the full list of median, interval range, and MLE values are shown in Table \ref{table:self-retrieval_best-fit_parameters}. As there is no cloud opacity in the simulated data, the model can adapt its cloud opacity in several ways to effectively remove its influence on the resulting emission spectrum. The first is to turn the reference optical depth $\tau\!\left(\lambda_0\right)$ to a very low value ($\ll 1$); the second is to introduce significant opacity $\tau\!\left(\lambda_0\right) \gtrsim 1$ but to place the cloud deep into the atmosphere, below where the majority of the thermal emission originates in the spectrum (i.e.~below the photosphere).}
\label{fig:self-retrieval_on-clear_cloud-corners}
\end{center}
\end{figure*}

The retrieval on simulated cloud-free data has no issue converging to an excellent fit, with the final $\chi_\nu^2$ value very close to 1 (Figure \ref{fig:self-retrieval_on-clear_spectrum} and Table \ref{table:self-retrieval_best-fit_parameters}). The retrieved C/O ratio is consistent with the input value of solar (0.54), with a 68\% confidence interval in the posterior distribution of $\pm 0.003$. The only abundance not tightly constrained is CO$_2$, with the true abundance sitting above the upper limit of the 68\% confidence interval. The retrieved T-P profiles show a weak constraint at either end of the pressure range, with pressures smaller than $\sim 10^{-3}$ or larger than $\sim 10$ bars. The best-fit T-P of the cloudy model is closer to the true profile at the shallowest pressures, but the confidence ranges of the two models overlap significantly at these pressures, meaning the relative fit qualities in this region of the atmosphere are not significantly different. The contribution plots show that most of the contribution is concentrated between pressures of a few tenths of a bar to a few bars; therefore it is not surprising that most of the uncertainty in the thermal profile arises away from these intermediate pressures.

The nominal expectation is that the cloudy model, when applied to cloud-free data, will effectively ``turn off'' the cloud opacity. This is largely true; the median opacity at the reference wavelength of 1 $\mu$m is very weak (an optical depth of $\sim 0.01$, yielding an attenuation of at most a few tenths of a percent). However, there is a tail in the distribution of optical depths; in some cases the model will choose non-negligible cloud opacity, but the increases in optical depth correlate with the depth of the cloud top. This is consistent with the relative lack of contribution to the emission spectra from pressures deeper than a few bars, which is where the distribution of optical depths reaches $\sim 1$. The single-scattering albedo $\omega_0$ is low, particularly in the low-opacity cases, and is weighted toward low power-law exponents, which would allow non-negligible cloud opacity at wavelengths $< 1$ $\mu$m. This may explain why the distribution of effective temperatures for the cloudy model fit peaks near the true value of 1821 K but has a substantial secondary peak, with the median at 1274 K. If one were to extend the forward models to shorter wavelengths, we would see these models diverge from their cloud-free counterparts. Regardless of the precise nature of the way in which the cloud model withholds its opacity from the spectrum, the retrieved distributions of the gas species are very similar, and the constraints on the C/O ratio are of nearly identical accuracy and precision. The cloud-free model returns a Bayes factor higher than its cloudy counterpart by a factor of approximately 18, meriting its preference following the interpretation of \citet{Jeffreys1998}. In the interpretation of \citet{Benneke2013}, we could say that we ``detect'' the cloud-free model with a significance slightly less than $3 \sigma$. The cloud-free model gains its advantage by virtue of using fewer free parameters to fit the data.

\subsection{Retrievals on Cloudy Data}\label{sec:self-results:results:power-law}
\begin{figure*}[htb]
\begin{center}
\textbf{Self-Retrieval on L Dwarf-like Cloudy Spectrum}
\includegraphics[width=17cm]{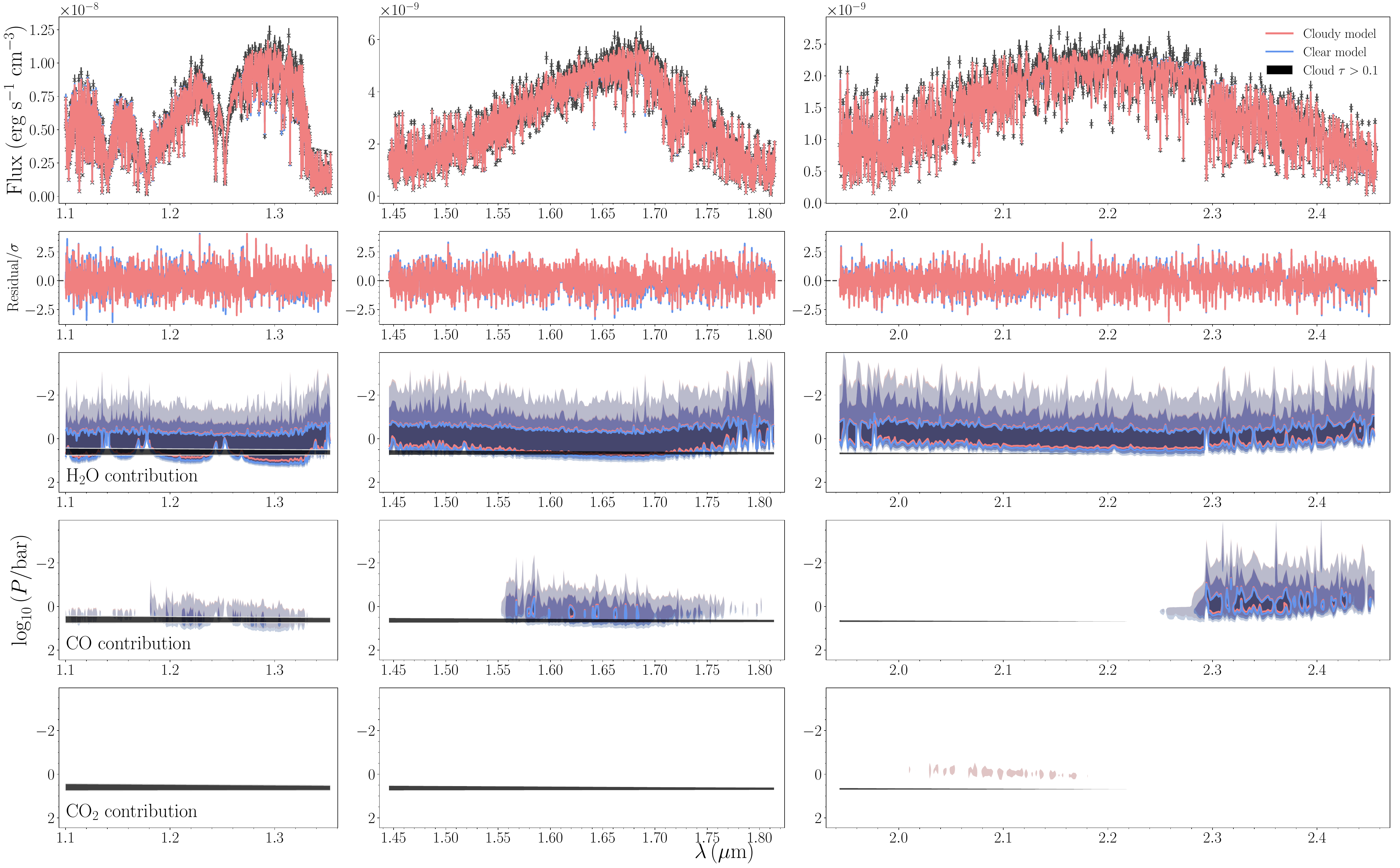}
\caption{Spectra and contribution functions of the retrieved forward model fits to data simulated from \textbf{cloudy} forward models of an L dwarf. The forward model spectra (using the MLE parameter values) are in color, with the data in grey. The retrieved models are nearly identical and overlap nearly entirely. Immediately beneath the spectra are the contribution functions for the 3 principal carbon and oxygen bearing species. The deepest contours, outlined in solid colors, enclose the regions where the contribution function reaches $>1$\% of the total contribution within the atmospheric column at a given wavelength bin. Each successive contour denotes 2 orders of magnitude smaller fractional contribution (here, $10^{-4}$ and $10^{-6}$).}
\label{fig:self-retrieval_on-cloudy_spectrum}
\end{center}
\end{figure*}

\begin{figure*}[htb]
\begin{center}
\textbf{Self-Retrieval on L Dwarf-like Cloudy Spectrum}\\
\includegraphics[width=13cm]{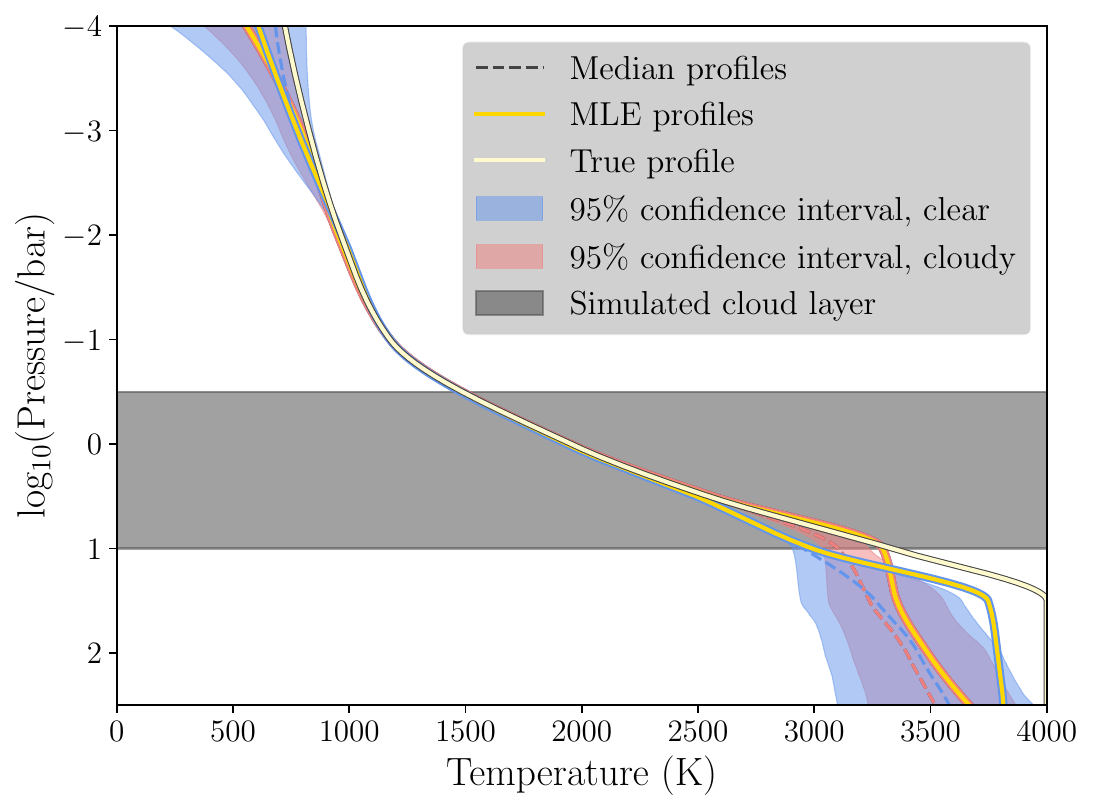}
\caption{The vertical temperature-pressure profiles of the retrieved forward model fits to data simulated from \textbf{cloudy} forward models of an L dwarf. We show the MLE, median, and 95\% confidence interval of the retrieved T-P profiles with the true profile over-plotted.}
\label{fig:self-retrieval_on-cloudy_T-P}
\end{center}
\end{figure*}

\begin{figure*}[htb]
\begin{center}
\textbf{Self-Retrieval on L Dwarf-like Cloudy Spectrum}
\includegraphics[width=17cm]{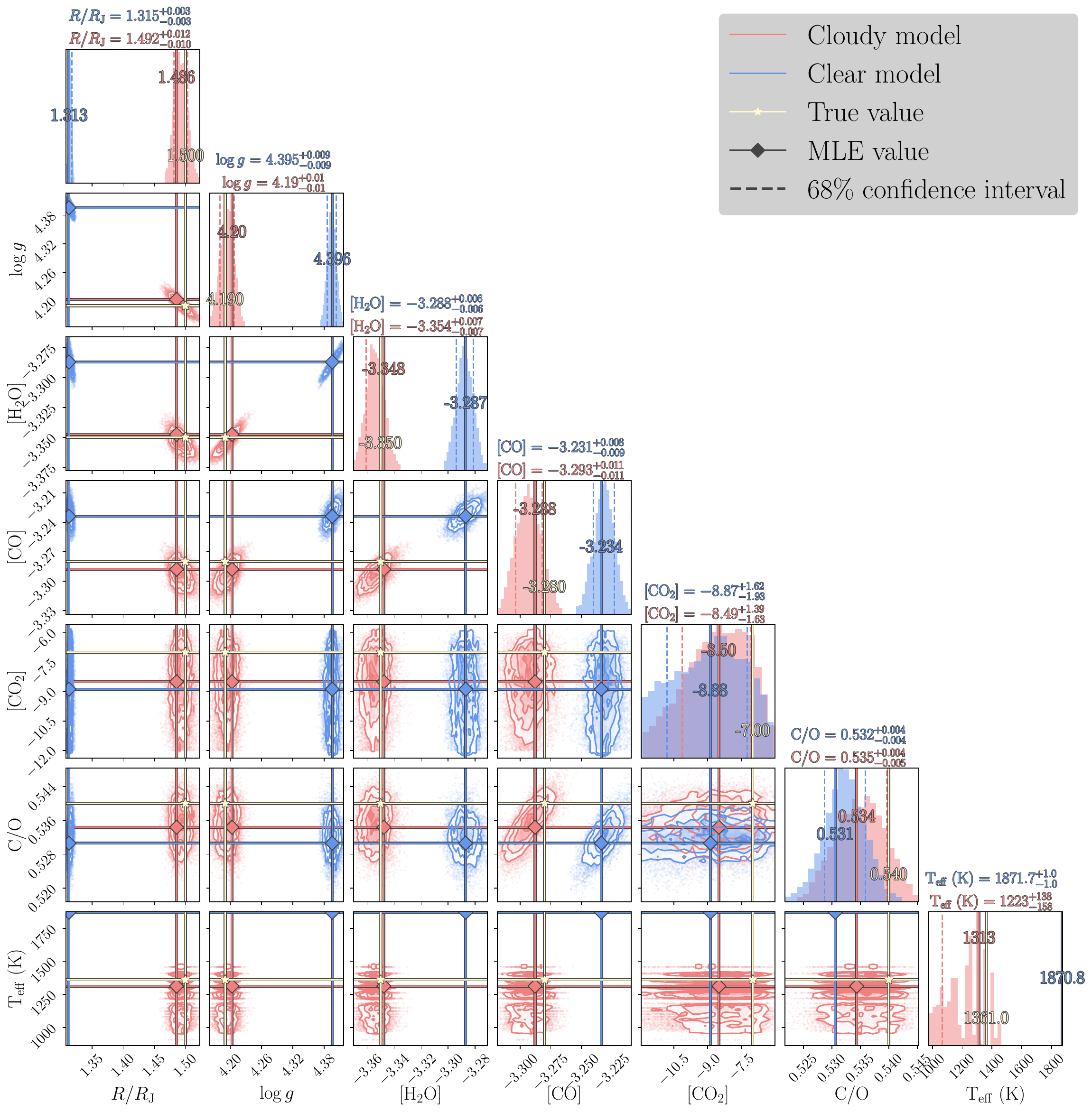}
\caption{A selection of parameters of the retrieved forward model fits to data simulated from \textbf{cloudy} forward models of an L dwarf, shown as 1-D and 2-D histograms in a corner plot of the retrieved posterior distributions. The median value and 68\% confidence interval of each parameter are shown at the top of each column; the full list of median, interval range, and MLE values are shown in Table \ref{table:self-retrieval_best-fit_parameters}.}
\label{fig:self-retrieval_on-cloudy_custom-corners}
\end{center}
\end{figure*}

\begin{figure*}[htb]
\begin{center}
\includegraphics[width=17cm]{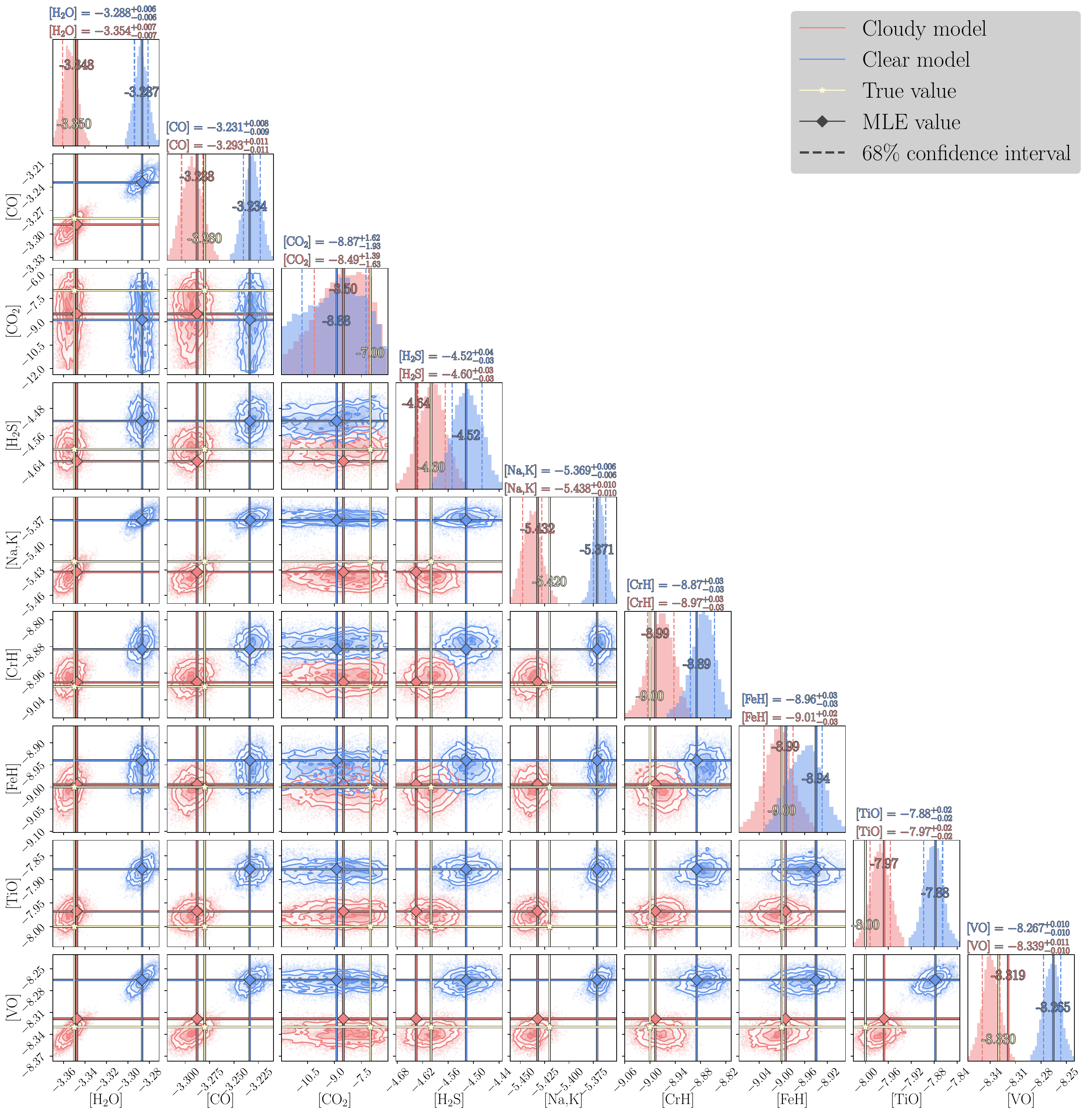}
\caption{Single-parameter (1-D) and parameter-versus-parameter (2-D) posterior distributions of gas parameters from the samples in the cloudy (red) and cloud-free (blue) forward model fit to data simulated from the cloudy forward model of an L dwarf (see \S \ref{sec:self-results}). The cloud opacity is modeled as a power law in wavelength, as described in \S \ref{sec:model:clouds}. The median value and 68\% confidence interval of each parameter are shown at the top of each column; the full list of median, interval range, and MLE values are shown in Table \ref{table:self-retrieval_best-fit_parameters}.}
\label{fig:self-retrieval_on-cloudy_gas-corners}
\end{center}
\end{figure*}

\begin{figure*}[htb]
\begin{center}
\includegraphics[width=17cm]{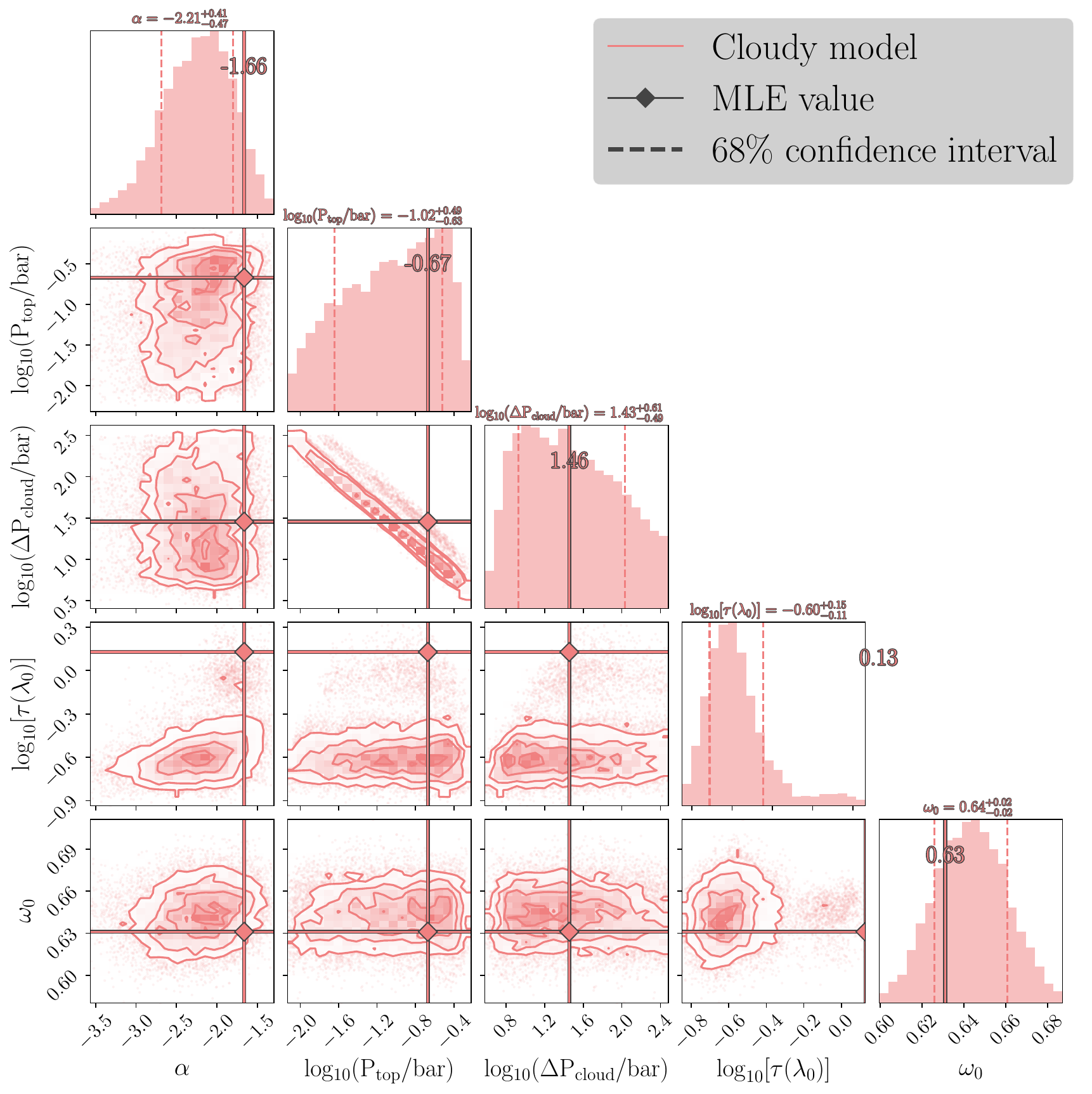}
\caption{Single-parameter (1-D) and parameter-versus-parameter (2-D) posterior distributions of cloud parameters from the samples in the cloudy (red) and cloud-free (blue) forward model fit to data simulated from the cloudy forward model of an L dwarf (see \S \ref{sec:self-results}). The cloud opacity is modeled as a power law in wavelength, as described in \S \ref{sec:model:clouds}. The median value and 68\% confidence interval of each parameter are shown at the top of each column; the full list of median, interval range, and MLE values are shown in Table \ref{table:self-retrieval_best-fit_parameters}.}
\label{fig:self-retrieval_on-cloudy_cloud-corners}
\end{center}
\end{figure*}

We now show the fits to mock data with clouds included, with the retrieved spectra and contributions in Figure \ref{fig:self-retrieval_on-cloudy_spectrum}, T-P profiles in Figure \ref{fig:self-retrieval_on-cloudy_T-P}, and posterior distributions of parameters in Figures \ref{fig:self-retrieval_on-cloudy_custom-corners}--\ref{fig:self-retrieval_on-cloudy_cloud-corners}. The power-law cloud opacity model yields a reduced chi-square statistic $\chi_\nu^2 \approx 1$. The retrieved C/O ratio is slightly less accurate when compared with that of the cloud-free case, with the true value lying just outside the 68\% confidence interval (but well within the 95\% interval). 

The cloud-free model applied to cloudy data returns a slightly worse fit, with the reduced chi-square statistic increasing to 1.07. In this case the model compensates for a lack of clouds by decreasing the radius, increasing the gravity, and increasing the abundances of all species except that of CO$_2$ by 0.05--0.1 dex. This allows for a C/O ratio distribution that is still marginally consistent with the truth at the 95\% confidence level. When comparing the quality of the fits between the cloudy and cloud-free models, we find a Bayes factor of approximately 150. This puts the modest difference in the reduced chi-square statistics in greater perspective; the cloud-free model is strongly disfavored when compared with the model with clouds, with a frequentist translation to a model which is preferred at $\sim 3.6 \sigma$. This is due to an accumulation of minor differences ($\lesssim$ the typical error bar) between the cloudy and clear fits, primarily in the $J$ band where the simulated cloud opacity is strongest.

While the cloudy model better fits the data, it does not reproduce all cloud properties precisely. The cloud top position and extent of the cloud are tightly correlated (see Figure \ref{fig:self-retrieval_on-cloudy_cloud-corners}), and the reference optical depth for the maximum likelihood estimator sits at the high tail of its posterior distribution, with the true value even farther out. The true value for the pressure of the cloud top sits at the high end of the 68\% confidence interval for the retrieved posterior distribution, which shows the model is able to reproduce where in the atmosphere cloud opacity should become significant. Since the reference optical depth is inaccurate, this suggests that there is some minimum opacity that effectively suppresses much of the emission deeper than the cloud; no additional opacity is needed, therefore the model finds a solution centered around the minimum sufficient opacity.

Finally, when comparing the retrieved T-P profiles, we see that, as in the test retrievals on spectra generated from a model without clouds, the weakest constraints in temperature arise in the shallowest and deepest parts of the atmosphere. However, we now see an additional divergence between the cloud-free and cloudy fits --- namely, the cloud-free profile diverges from the true profile within the simulated cloud layer, while the cloudy profile remains close to the true profile. Then, by the time we reach the deepest extent of the clouds, both profiles have begun to diverge from the true profile. This suggests that the cloud-free model is attempting to compensate for its lack of clouds by keeping the temperature gradient shallower, suppressing its thermal contribution in a way that can mimic the effect of a cloud layer.

\subsection{Lessons from Self-Retrievals}\label{sec:self-results:lessons}
Taking the results of self-retrievals on cloudy and clear simulated data together, we demonstrate that our code is able to both identify the correct abundances and thermal structure in the photosphere in an atmospheric simulation. The results also highlight what we might not expect to constrain precisely due to theoretical limitations, such as the deepest parts of the T-P profile, and the opacity profile of the clouds. Additionally, a cloud-free model may be able to reproduce a C/O ratio consistent with the true value, but risks returning an inaccurate radius and gravity, molecular abundances that are almost all consistently too high, all with a temperature profile that is consistent at the same pressure ranges as the cloudy model, but with higher uncertainties at the lowest pressures. Additionally, when clouds are present, we expect a cloud-free model to show the greatest difference from that of a cloudy model within the cloud layer itself, changing its gradient to compensate for the lack of extinction from condensates. The bias in molecular abundances suggests that, at least in this wavelength range, the shape of the spectrum is determined more by the relative abundances than by the absolute abundances; put another way, we may expect to see a potential degeneracy between the T-P profile, gravity, and key molecular abundances, but nevertheless may expect the retrieved C/O ratio to not be significantly biased away from the true value. However, these conclusions are necessarily limited to which physics we choose to include in the model used to simulate the data; we are limited to commenting on the efficacy of the code in terms of the consistency of retrievals with the assumptions we have made.

\section{Retrieval on a Previously Characterized L Dwarf}\label{sec:test-results}
\begin{figure*}[htb]
\begin{center}
\textbf{Retrieval on 2M2224-0158 Spectrum}\\
\includegraphics[width=17cm]{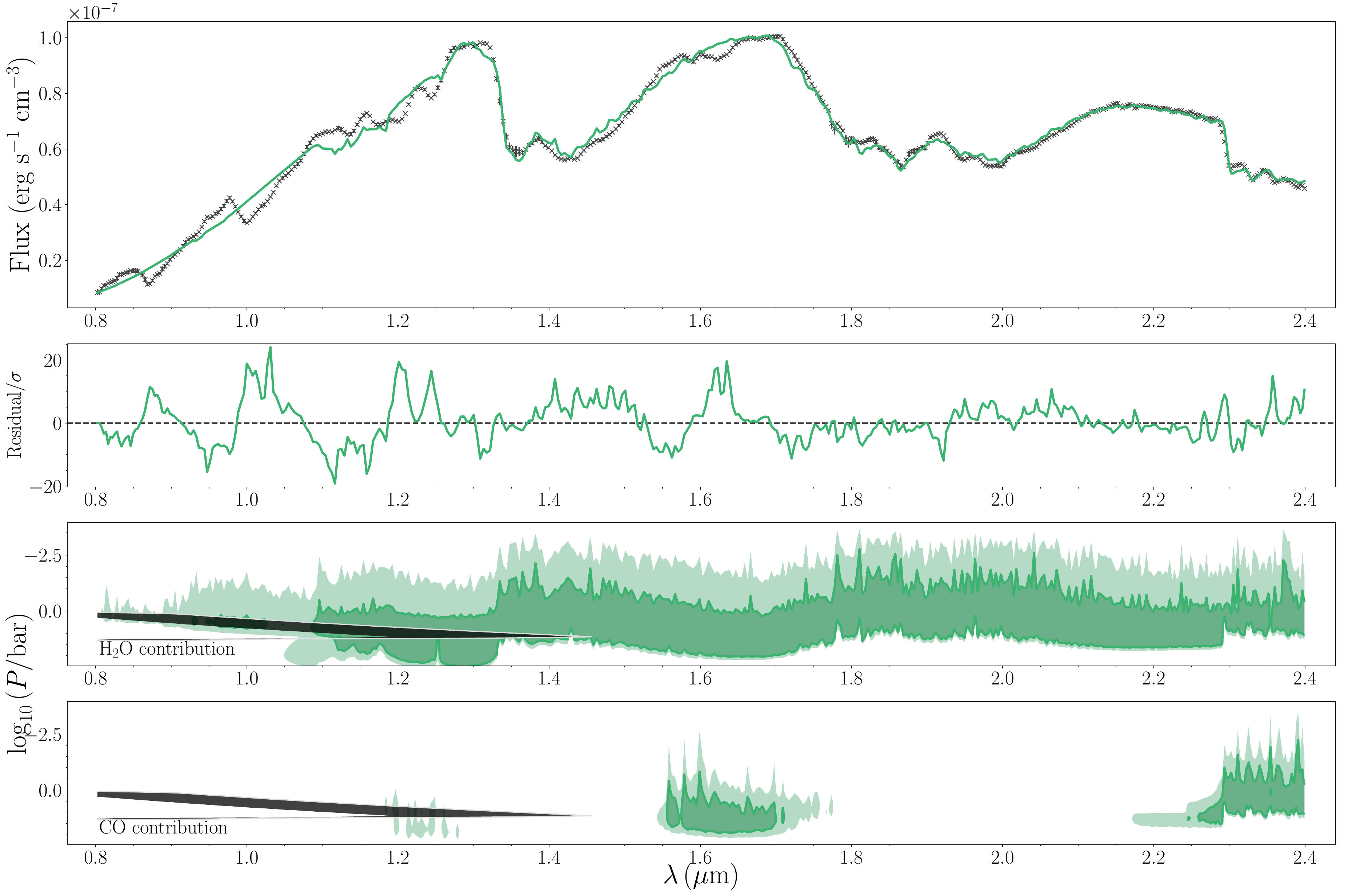}
\caption{Spectra and contribution functions of the retrieved forward model fit to the SpeX data for 2M2224. The forward model spectra (using the MLE parameter values) are in color, with the data in grey. Immediately beneath the spectra are the contribution functions for the 2 principal carbon and oxygen bearing species; CO$_2$ is not included in the retrieval as it was excluded from the retrieval in \citealt{Burningham2017}, inviting a more direct comparison of our results. The deepest contours, outlined in solid colors, enclose the regions where the contribution function reaches $>1$\% of the total contribution within the atmospheric column at a given wavelength bin. Each successive contour denotes 2 orders of magnitude smaller fractional contribution (here, $10^{-4}$).}
\label{fig:2M2224_spectrum}
\end{center}
\end{figure*}

\begin{figure*}[htb]
\begin{center}
\textbf{Retrieval on 2M2224-0158 Spectrum}\\
\includegraphics[width=13cm]{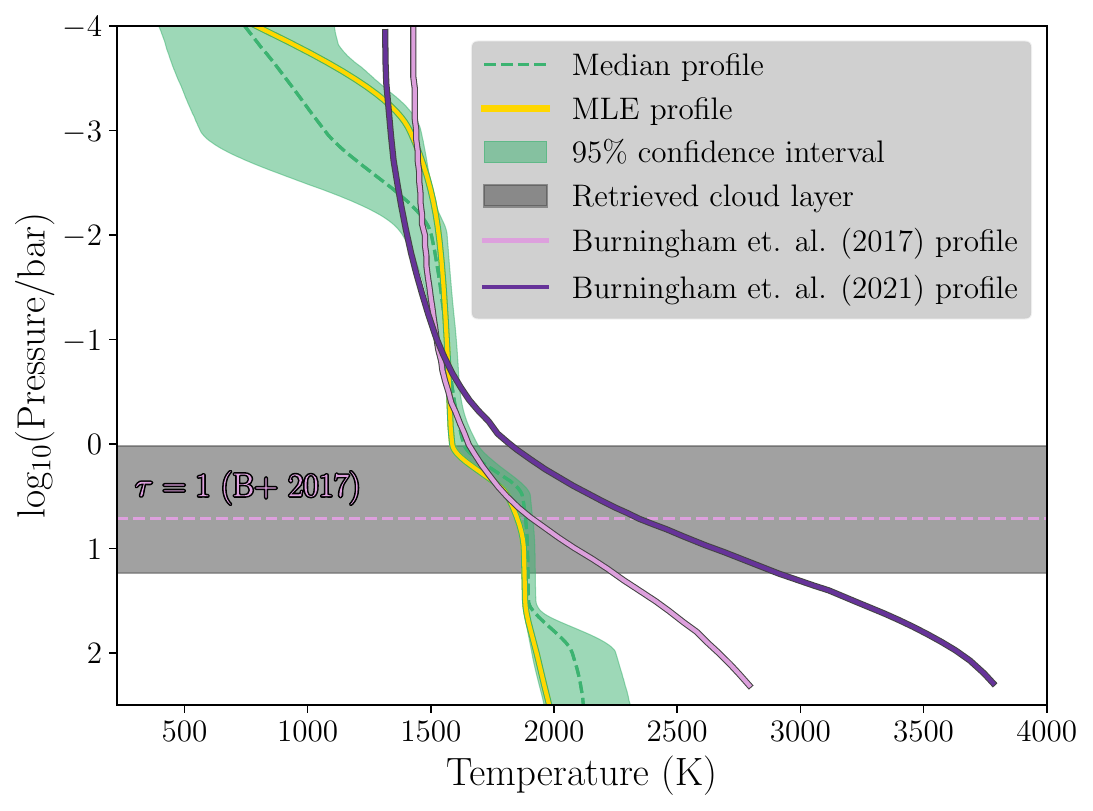}
\caption{The vertical temperature-pressure profiles of the retrieved forward model fit to the SpeX data for 2M2224. We show the MLE, median, and 95\% confidence interval of the retrieved T-P profiles, with the retrieved T-P profiles of retrievals from \citet{Burningham2017} and \citet{Burningham2021} also plotted for comparison. The latter profile is shown to highlight how the retrieved vertical structure changes as longer wavelength data are included. The median retrieved $\tau=1$ pressure for the retrieved deck cloud models of \citet{Burningham2017} is shown as a dashed line.}
\label{fig:2M2224_T-P}
\end{center}
\end{figure*}

\begin{figure*}[htb]
\begin{center}
\textbf{Retrieval on 2M2224-0158 Spectrum}\\
\includegraphics[width=17cm]{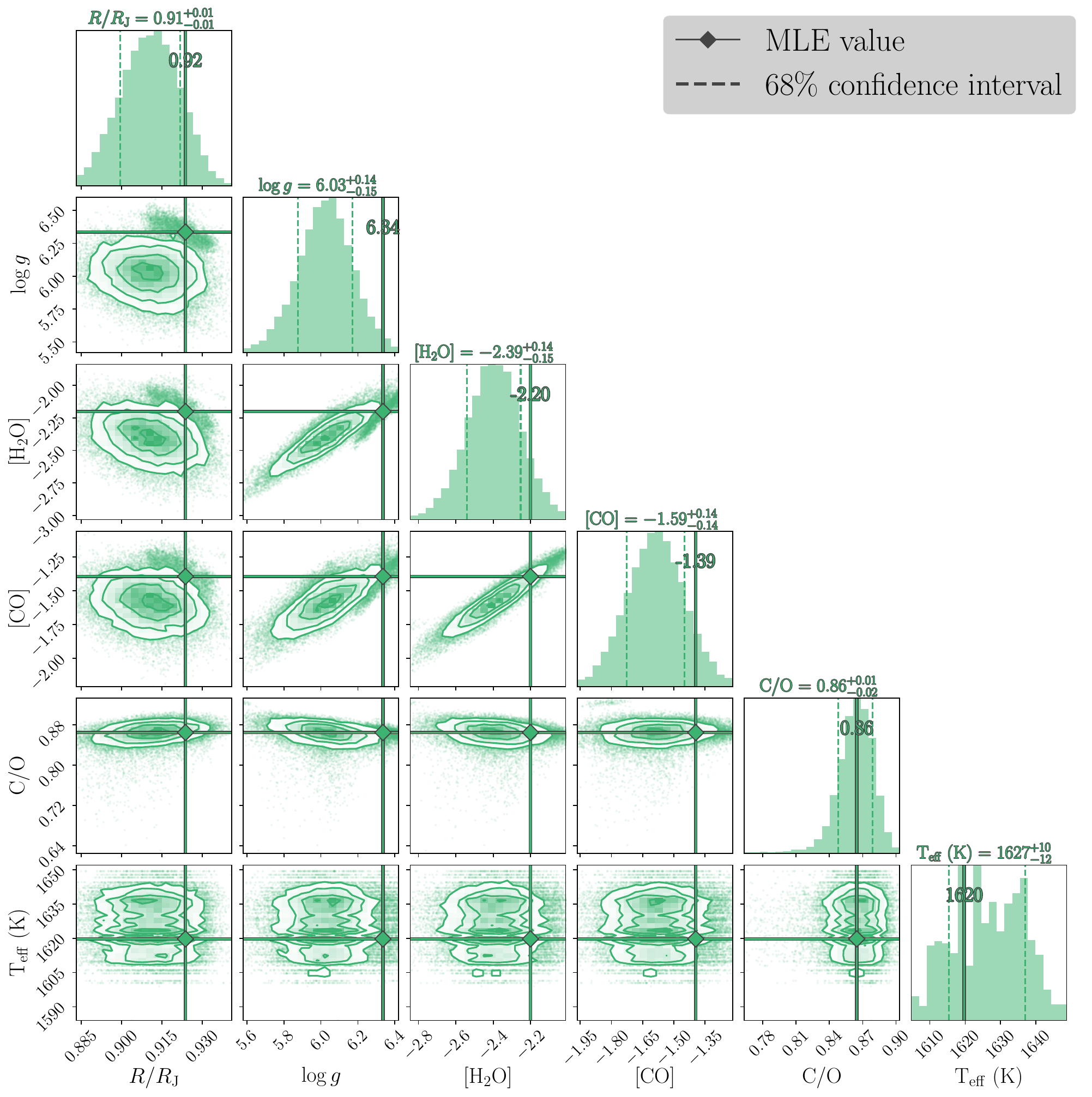}
\caption{A selection of parameters of the retrieved forward model fit to the SpeX data for 2M2224, shown as 1-D and 2-D histograms in a corner plot of the retrieved posterior distributions for selected parameters. The median value and 68\% confidence interval of each parameter are shown at the top of each column; the full list of median, interval range, and MLE values are shown in Table \ref{table:2M2224_best-fit_parameters}. Median values for the equivalent parameters in the run in \citet{Burningham2017} are listed in Table \ref{table:2M2224_best-fit_parameters}.}
\label{fig:2M2224_custom-corners}
\end{center}
\end{figure*}

\begin{figure*}[htb]
\begin{center}
\textbf{Retrieval on 2M2224-0158 Spectrum}\\
\includegraphics[width=17cm]{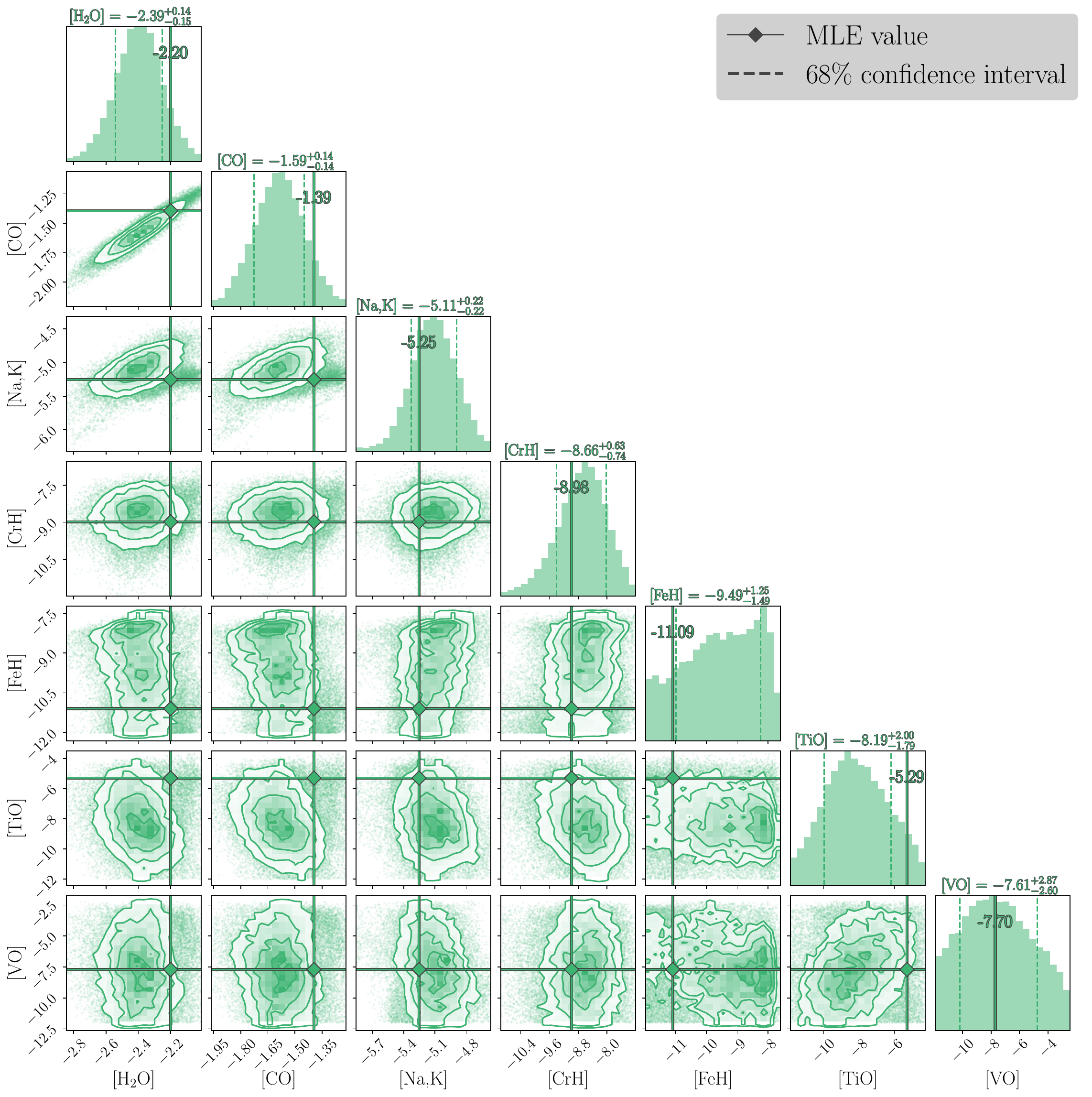}
\caption{Single-parameter (1-D) and parameter-versus-parameter (2-D) posterior distributions of gas parameters from the samples in the model fit to the SpeX data for 2M2224. The cloud opacity is modeled as a power law in wavelength, as described in \S \ref{sec:model:clouds}. The median value and 68\% confidence interval of each parameter are shown at the top of each column; the full list of median, interval range, and MLE values are shown in Table \ref{table:2M2224_best-fit_parameters}. Median values for the equivalent parameters in the run in \citet{Burningham2017} are listed in Table \ref{table:2M2224_best-fit_parameters}.}
\label{fig:2M2224_gas-corners}
\end{center}
\end{figure*}

\begin{figure*}[htb]
\begin{center}
\textbf{Retrieval on 2M2224-0158 Spectrum}\\
\includegraphics[width=17cm]{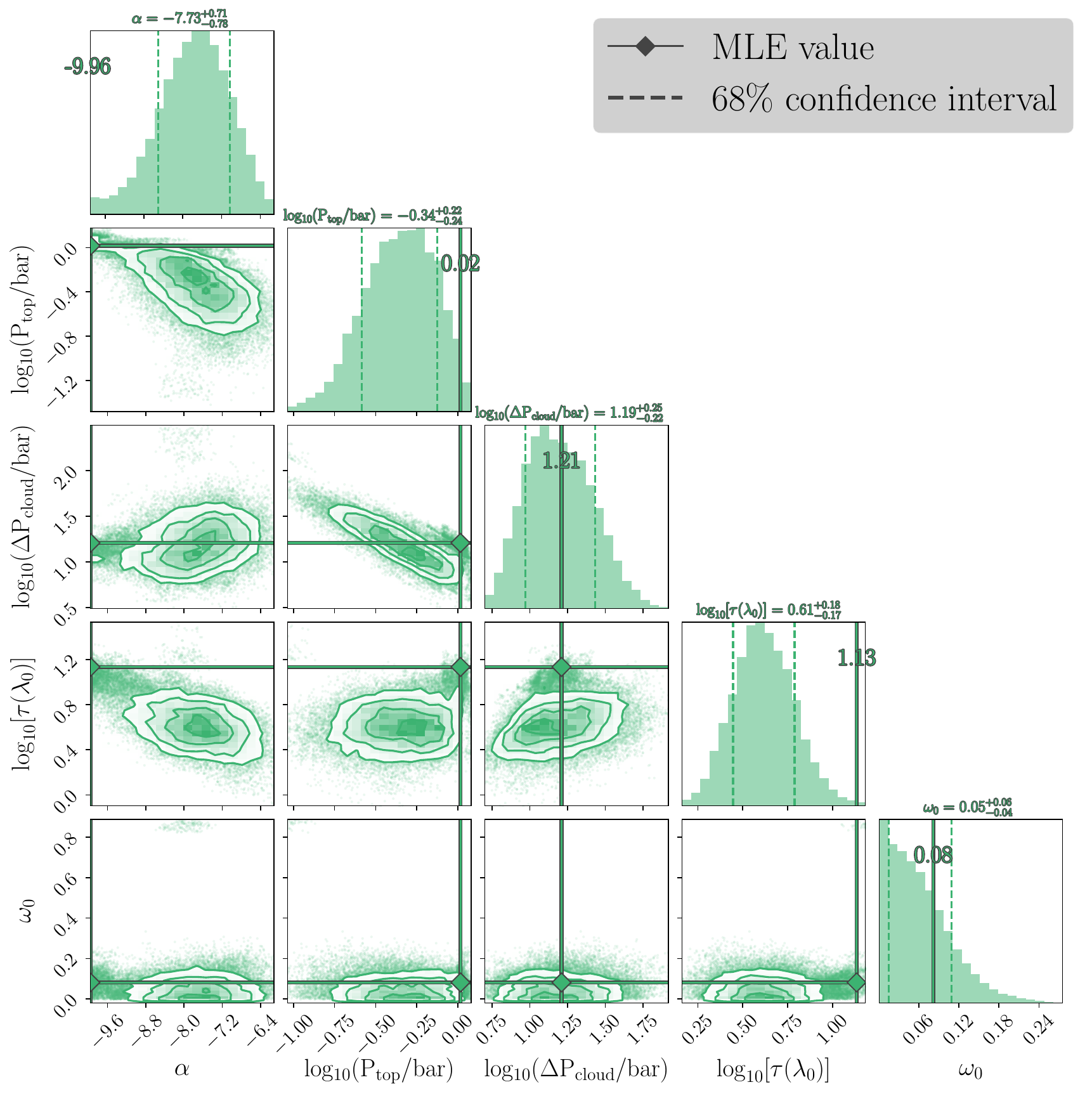}
\caption{Single-parameter (1-D) and parameter-versus-parameter (2-D) posterior distributions of cloud parameters from the samples in the model fit to the SpeX data for 2M2224. The cloud opacity is modeled as a power law in wavelength, as described in \S \ref{sec:model:clouds}. The median value and 68\% confidence interval of each parameter are shown at the top of each column; the full list of median, interval range, and MLE values are shown in Table \ref{table:2M2224_best-fit_parameters}. Median values for the equivalent parameters in the run in \citet{Burningham2017} are listed in Table \ref{table:2M2224_best-fit_parameters}.}
\label{fig:2M2224_cloud-corners}
\end{center}
\end{figure*}

\begin{deluxetable}{lccc}
\tabletypesize{\scriptsize}
\tablewidth{0pt}
\tablecaption{Median and MLE parameter values for the retrieval on 2M2224, as described in \S \ref{sec:test-results}, as well as the ranges of retrieved parameters reported in \citet{Burningham2017}. \label{table:2M2224_best-fit_parameters}}
\tablehead{
           \colhead{Name} &
           \colhead{Median} &
           \colhead{MLE} & 
           \colhead{Median \citep{Burningham2017}}
           }
\startdata
\cutinhead{Fit Quality}
$\chi_\nu^2$ & & 42 \\
\cutinhead{Fundamental}
$R/R_\mathrm{J}$                                              & $0.91\pm0.01$ & 0.92 & $0.93\pm0.03$ \\
$\log_{10}\!\left[g/\!\left(\mathrm{cm~s}^{-2}\right)\right]$ & $6.03^{+0.14}_{-0.15}$ & 6.34 & $5.31^{+0.04}_{-0.08}$ \\
$M/M_\mathrm{J}$                                              & $354^{+140}_{-103}$  & 744 & $72.22^{+6.25}_{-12.05}$ \\
$T_\mathrm{eff}$ (K)                                          & $1627^{+10}_{-12}$ & 1620 & $1723.34^{+18.03}_{-18.91}$ \\
C/O                                                           & $0.86^{+0.01}_{-0.02}$ & 0.86 & $0.85^{+0.06}_{-0.08}$ \\
Metallicity                                                   & $1.61\pm0.14$ & 1.80 & \tablenotemark{a}\\
\cutinhead{Gases ($\log_{10}$ number abundance relative to total)}
H$_2$O   & $-2.39^{+0.14}_{-0.15}$  &  $-2.20$ & $-3.16^{+0.08}_{-0.07}$ \\
CO       & $-1.59\pm0.14$           &  $-1.39$ & $-2.40^{+0.16}_{-0.14}$ \\
Na+K     & $-5.11\pm0.22$           &  $-5.25$ & $-5.33^{+0.23}_{-0.35}$ \\
CrH      & $-8.66^{+0.63}_{-0.74}$  &  $-8.98$ & $-7.49^{+0.25}_{-0.20}$ \\
FeH      & $-9.49^{+1.25}_{-1.49}$  & $-11.09$ & $-7.71^{+0.09}_{-0.12}$ \\
TiO      & $-8.19^{+2.00}_{-1.79}$  &  $-5.29$ & $-8.60^{+0.93}_{-2.19}$ \\
VO       & $-7.61^{+2.87}_{-2.60}$  &  $-7.70$ & $-9.59^{+0.83}_{-1.44}$ \\
\cutinhead{Temperature-Pressure\tablenotemark{b}}
$T_{-4}$  (K) & $715^{+204}_{-190}$   & 790  & \\
$T_{-3}$  (K) & $977^{+204}_{-218}$   & 1414 & \\
$T_{-2}$  (K) & $1500^{+39}_{-56}$    & 1532 & \\
$T_{-1}$  (K) & $1563^{+22}_{-24}$    & 1565 & \\
$T_{0}$   (K) & $1638\pm25$           & 1587 & \\
$T_{0.5}$ (K) & $1871^{+20}_{-25}$    & 1813 & \\
$T_{1}$   (K) & $1897^{+13}_{-12}$    & 1877 & \\
$T_{1.5}$ (K) & $1901^{+13}_{-12}$    & 1883 & \\
$T_{2}$   (K) & $2064^{+88}_{-66}$    & 1929 & \\
$T_{2.5}$ (K) & $2103^{+97}_{-74}$    & 1981 & \\
\cutinhead{Clouds}
$\alpha$                                                         & $-7.73^{+0.71}_{-0.78}$   & $-9.96$ & $-2.66^{+0.63}_{-1.45}$ \\
$\log_{10}\!\left( P_\mathrm{top}/\mathrm{bar} \right)$          & $-0.34^{+0.22}_{-0.24}$   &  $0.02$ & $0.71^{+0.10}_{-0.06}$ \\
$\log_{10}\!\left( \Delta P_\mathrm{cloud}/\mathrm{bar} \right)$ & $1.19^{+0.25}_{-0.22}$    &   1.21  & $3.69^{+2.28}_{-2.38}$ \\
$\log_{10}\!\left[\tau\!\left(\lambda_0\right)\right]$           & $0.61^{+0.18}_{-0.17}$    &   1.13  & \tablenotemark{a} \\
$\omega_0$                                                       & $0.05^{+0.06}_{-0.04}$    &   0.08  & $0.52^{+0.22}_{-0.29}$ \\
\enddata
\tablenotetext{a}{Bulk metallicity and reference optical depth not reported.}
\tablenotetext{b}{We use a different T-P parametrization from that of \citet{Burningham2017}. A plot comparing the two retrieved profiles is available in Figure \ref{fig:2M2224_T-P}.}
\end{deluxetable}

Our first true retrieval is of the mid-L field dwarf 2MASSW J2224438--015852 \citep{Kirkpatrick2000}, which we refer to here as 2M2224. This is one of the brown dwarfs studied with the \textit{Brewster} retrieval code in \citet{Burningham2017}, and with mid-infrared data in \citet{Burningham2021}. To benchmark our code against previously published results in the \JHK{} spectral range, we limit our re-analysis to the original, $R \sim 75$ spectrum from \citet{Burgasser2010}\footnote{The reduced spectrum is available at \url{http://pono.ucsd.edu/~adam/browndwarfs/spexprism/html/ldwarf.html}.}. The full wavelength range of the data is 0.65--2.56 $\mu$m, but to compare with \citet{Burningham2017} we choose to only use the range from 0.8--2.4 $\mu$m, though a retrieval was performed with the full dataset. \citet{Burningham2017} retrieve an effective temperature $T_\mathrm{eff} = 1723^{+18}_{-19}$ K and $\log g = 5.31^{+0.04}_{-0.08}$. From the retrieved distributions of their H$_2$O and CO abundances, we infer a C/O ratio of 0.85$^{+0.06}_{-0.08}$. Our forward model follows nearly the same parametrization, with a few differences: our opacities lack CaH, and we use the \citet{Piette2020} T-P profile parametrization, which can reproduce the same shapes as the Madhusudan-Seager model but is slightly more flexible. Our cloud model is functionally equivalent to the ``slab'' case as described in \S 2.1.3 of \citet{Burningham2017}, though we choose to include the reference optical depth ($\tau_0 \equiv \tau\!\left(\lambda=1\,\mu\mathrm{m}\right)$) as a free parameter in our corner plots. It should be noted, however, that \citet{Burningham2017} report their results using a ``deck'' cloud model, though both that and the slab model were tested in their work. Finally, our sampling method differs in that \citet{Burningham2017} used an MCMC parameter estimation technique, and imposed more restrictive priors on gravity to keep their mass below the nominal main-sequence limit of 80 $M_\mathrm{J}$. Our choices of a more free T-P profile parametrization and wider priors on the gravity mean that our code can explore a broader range of solutions for the vertical atmospheric structure, but with a concession that our solution has a higher risk of introducing structure to the profile that does not have a feasible physical interpretation.

Results from the retrieval are shown in Figure \ref{fig:2M2224_spectrum} for the spectrum and contributions, Figure \ref{fig:2M2224_T-P} for the retrieved T-P profiles, and posterior distributions for parameters in Figures \ref{fig:2M2224_custom-corners}--\ref{fig:2M2224_cloud-corners}. Our retrieved spectrum does not precisely reproduce the shapes of the local peaks in the $J$ and $H$ bands, and generally prefers a ``smoother'' (though not necessarily better) fit to the spectrum than that retrieved in Figure 8 of \citet{Burningham2017}. Our retrieved C/O ratio of $0.86^{+0.01}_{-0.02}$ sits entirely within the confidence interval reported in \citet{Burningham2017}, despite retrieving a higher gravity and higher abundances particularly in H$_2$O and CO. Our model finds a solution that prefers a higher metallicity ($1.61\pm0.14$),  Our T-P profile mimics the shape of the 2017 paper from $\sim 0.01$ bars until the location of our retrieved cloud layer, where the models then diverge. The T-P profiles diverge most strongly where the preferred deck cloud model of \citet{Burningham2017} reaches an optical depth of 1 (at $\log_{10}\!\left(P/\mathrm{bar}\right)=0.71$). The extent of our cloud layer encompasses their median $\tau=1$ pressure, but our retrieved power-law dependence in wavelength is much more steeply negative than theirs, with a median $\alpha=-7.73^{+0.71}_{-0.78}$. In other words, our model prefers a solution that allows significant cloud opacity in the $J$-band portion of the spectrum, but rapidly diminishes at longer wavelengths. Our model has difficulty finding a solution to the atmosphere from $\approx 0.8$--1.3 $\mu$m, given that the model fit fails to capture the smaller-scale variations in the data in the region where it determines clouds are most significant. This may mean that the cloud model is instead being used to compensate for an inability to fit this portion of the spectrum, while fitting the remainder in the $H$- and $K_\mathrm{s}$-band ranges more accurately. An earlier retrieval with the full 0.65--2.56 $mu$m dataset did return a cloud power-law exponent of $\alpha = -2.04\pm0.02$, but also had its own difficulties in capturing the entire spectrum, with a better fit to the 0.8--1.2 $\mu$m region but a worse fit in the $K_\mathrm{s}$ band from 2.00--2.35 $\mu$m, and a similarly high gravity.

This agreement in C/O despite disagreement elsewhere is similar to the findings in works such as \citet{Molliere2020}, where their tests of models with different cloud models yielded similar C/O ratios despite retrieving disagreeing thermal and cloud profiles. In our case, with the more flexible thermal structure, there is an additional degeneracy between the gravity and the molecular abundances/metallicity. The higher the metallicity, the less deep in the atmosphere a given optical depth will be reached, but the higher the gravity, the smaller a path length for a given change in pressure, meaning that the equivalent optical depth will occur at a higher pressure. Our choice of T-P profile allows flexibility in adapting the shape of the vertical thermal profile to changes in model gravity and metallicity; therefore, we expect gravity and metallicity to be negatively correlated. Mirroring the behavior we saw in the cloud-free versus cloudy models applied to simulated data in \S \ref{sec:self-results}, it is possible to retrieve an accurate C/O ratio by retrieving abundances that are accurate relative to each other, but biased in their absolute values. It is difficult to compare the shallow thermal gradient with the behavior suggested in \citet{Tremblin2016}, where a shallow temperature gradient driven by a thermo-chemical instability can mimic some of the spectral behavior attributed to clouds, since in this case both the shallow gradient and significant cloud opacity are present in the model solution. Nevertheless, we keep these findings in mind when interpreting the results of our retrieval on HD 106906 b (\S \ref{sec:results}).

\section{Retrieved Atmospheric Properties for HD 106906 b}\label{sec:results}
\subsection{Retrieval Setup: Single-Band Trials and Regions of High Tellurics}\label{sec:results:setup}
\begin{figure*}[htb]
\begin{center}
\begin{tabular}{ccc}
\includegraphics[width=5.5cm]{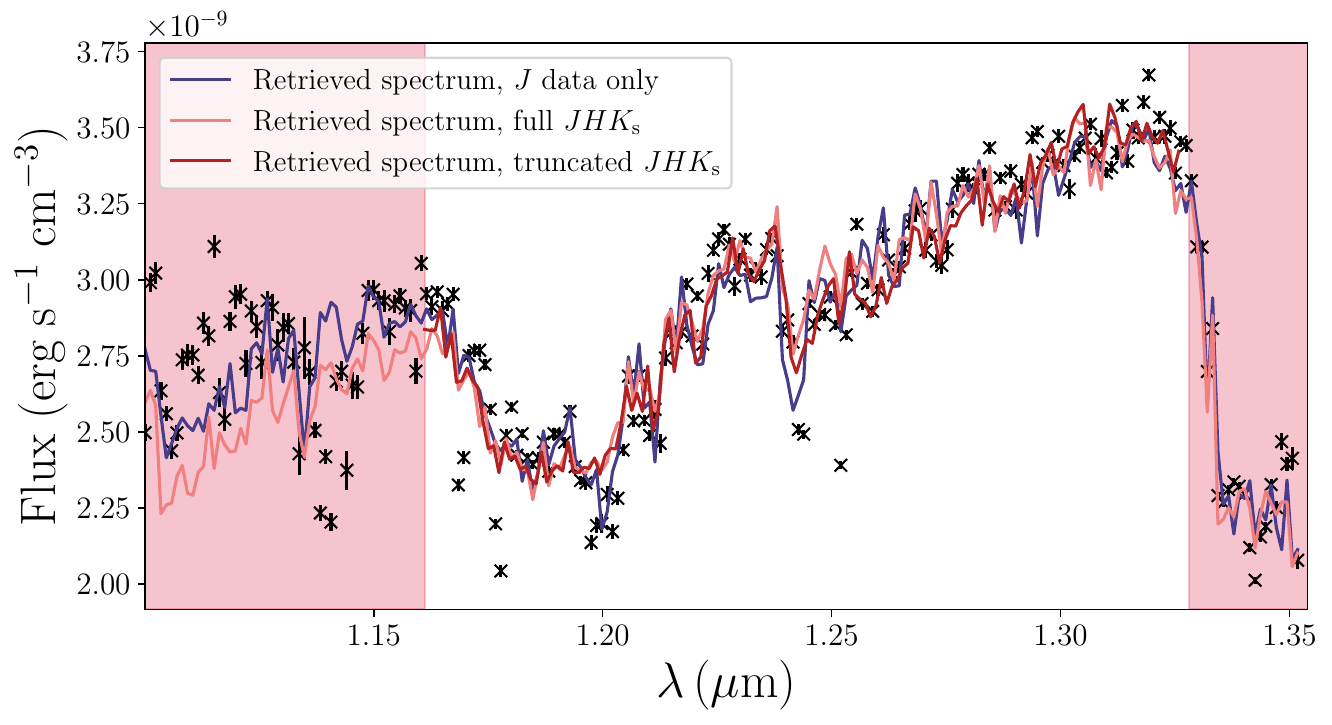} & \includegraphics[width=5.5cm]{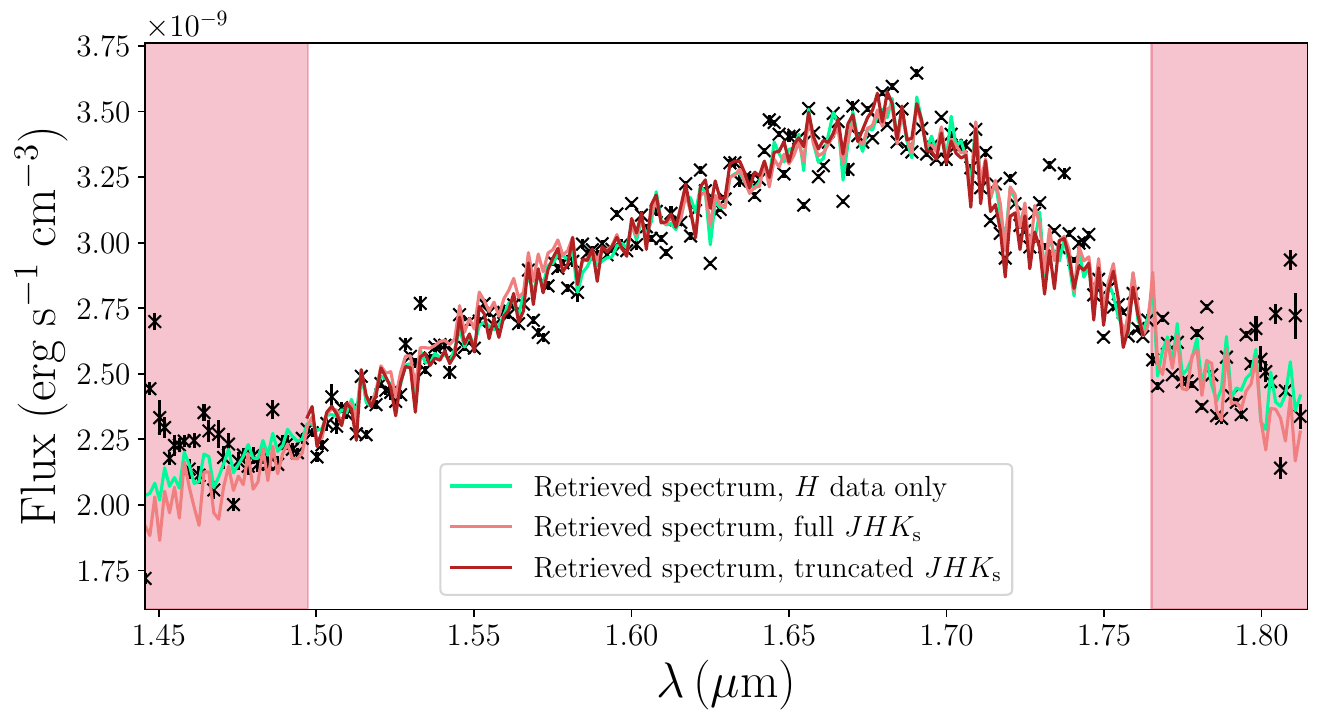} & \includegraphics[width=5.5cm]{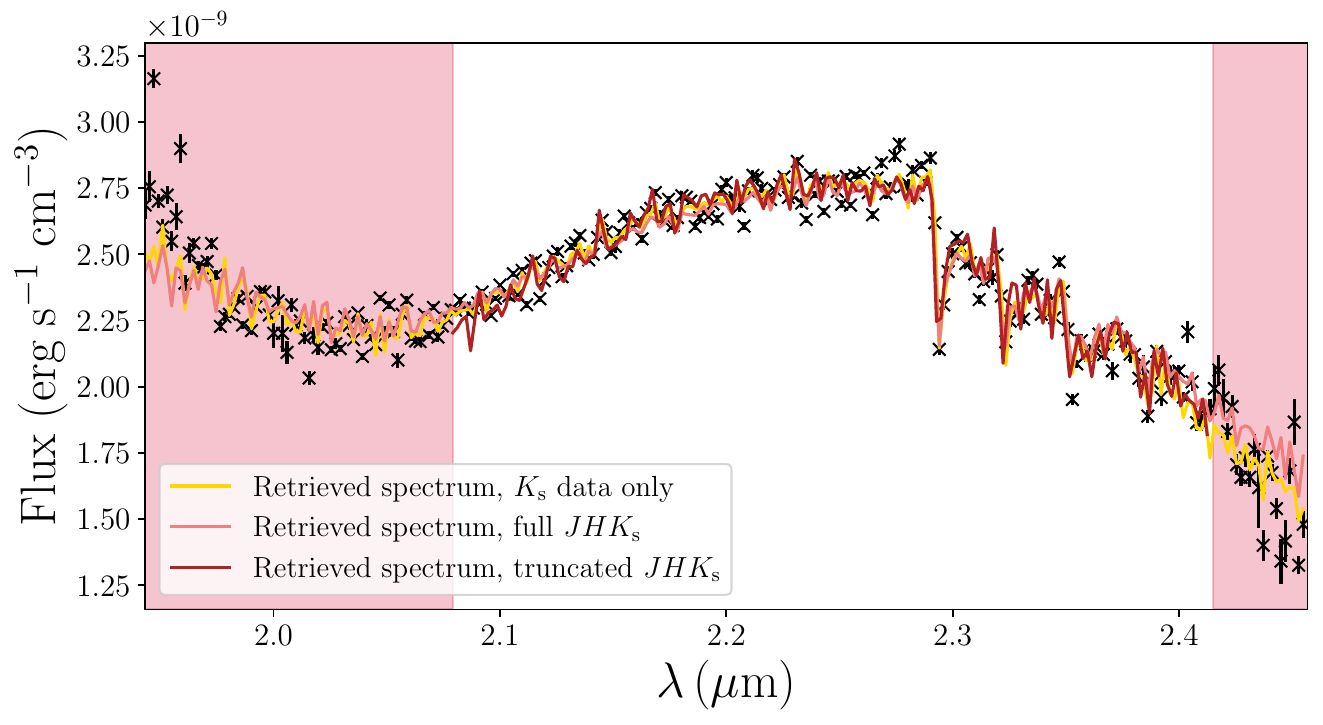} \\
\end{tabular}
\caption{Initial retrieval fits to the HD 106906 b spectrum. The data are down-sampled to a maximum resolution of $\approx 500$, a factor of 4--8 lower than the original. The fit using the full dataset is shown in light red. The shaded pink regions indicate where \citet{Daemgen2017} identified contiguous or near-contiguous regions of suspected high telluric contamination that could not be reliably fully removed in reduction, thus introducing potential residual systematics. The fits with each band individually are shown in the various non-red colors in each band (purple for $J$, green for $H$, and yellow for $K_\mathrm{s}$). The fit using the data across all bands, but without the high-telluric regions, is shown in darker red. Both the full and single-band fits perform most poorly in fitting the data in the high-telluric regions, especially on the blue ends of each band.}
\label{fig:HD106906b_PLO_initial-fits}
\end{center}
\end{figure*}

When moving from the test retrievals on simulated data to retrievals on the actual HD 106906 b data, there are a few differences in the model setup, though the core physical model remains the same. The first is the addition of calibration terms that scale the flux in each band by a multiplicative constant; this is to account for uncertainties in the photometry, as discussed in \S \ref{sec:data:data}. These calibration scales are partly degenerate with the retrieved radius, so when reporting these calibration scales, we normalize the radius such that in each case, the effective calibration scale in the $K_\mathrm{s}$ band is 1. The second is to add a parameter for fractional cloud cover ($f_\mathrm{cloud}$). Since we are using a 1-D (vertical-only) model, the fractional cloud cover is assumed to be isotropic, and the emission flux is simply weighted between the fully-cloudy flux one calculates from the given parameters ($F_\mathrm{cloudy}$) and the flux given the same parameters but without clouds ($F_\mathrm{clear}$):
\begin{equation}
    F = f_\mathrm{cloud} F_\mathrm{cloudy} + \left(1-f_\mathrm{cloud}\right) F_\mathrm{clear}.
\end{equation}
The final modification is that the data are down-sampled to a maximum resolution of $\approx 500$, a factor of 4--8 lower than the original. This is because the maximum resolution of our opacity tables is 50,000, and to avoid introducing excess artificial noise from binning effects, we impose a limit of $R_\mathrm{opacity}/R_\mathrm{data} \geq 100$, which requires us to re-sample the data to a lower resolution. We calculate the uncertainties in the down-sampled data as uncertainties in the mean, i.e.~since we now have $N=4$--8 resolution elements of the original spectrum in each of the down-sampled elements, our uncertainties in each new element are assumed to be smaller by a factor of $\sqrt{N-1}$. This is a lower limit of the true uncertainties in the new spectrum, as the errors between the original pixels and therefore resolution elements are almost certainly correlated at some level. To make an estimate of the typical correlation length, we use the approach in \S 2.2 of \citet{Line2015}, where one calculates the auto-correlation of the residuals for an initial model fit to the data. Doing this, we find that the auto-correlation drops and subsequently remains at or below $\approx 0.25$ at a scale of 6--8 pixels. Therefore, our reported fit qualities, such as chi-squared statistics, may be overestimated by roughly a factor of 2--3. However, a larger source of systematic errors comes in the form of telluric contamination, which is typically strongest at the boundaries of each band. These errors are more likely to introduce biases in the retrieved atmospheric parameters, and so we perform retrievals to examine the effects of including versus excluding these portions of the spectrum.

An initial retrieval was run on the full HD 106906 b dataset, along with retrievals on data from each band ($J$, $H$, and $K_\mathrm{s}$) individually; see Figure \ref{fig:HD106906b_PLO_initial-fits}. The purpose of this initial set of runs was to understand how well the retrieval could fit the data, and, depending on whether and how the individual band fits compare with the full-spectrum fit, suggest whether the model is capturing the physics of the companion's atmosphere with adequate flexibility (for example, this comparison can provide a first-order check to whether the assumption of constant mixing ratios is sufficient). The result is broadly that there are wavelength regions of each band where \emph{neither} the single-band nor the full-spectrum fits the data well. The largest discrepancy occurs at the blue end of each band, as well as to a lesser extent at the red end of the $H$ band. These regions are consistent with the wavelength regions identified in the original publication of the data \citep[][specifically Figure 2]{Daemgen2017}, where telluric contamination is thought to affect the data reduction most severely. Given the overlap of the most poorly fit regions with the suspected high-telluric regions, we choose to excise these data from the final retrievals. The final fit, using the truncated data across all bands, is over-plotted in Figure \ref{fig:HD106906b_PLO_initial-fits}.

\subsection{The Cloudy Model}\label{sec:results:power-law}
\begin{figure*}[htb]
\begin{center}
\textbf{Retrieval on HD 106906 b Spectrum} \\
\includegraphics[width=17cm]{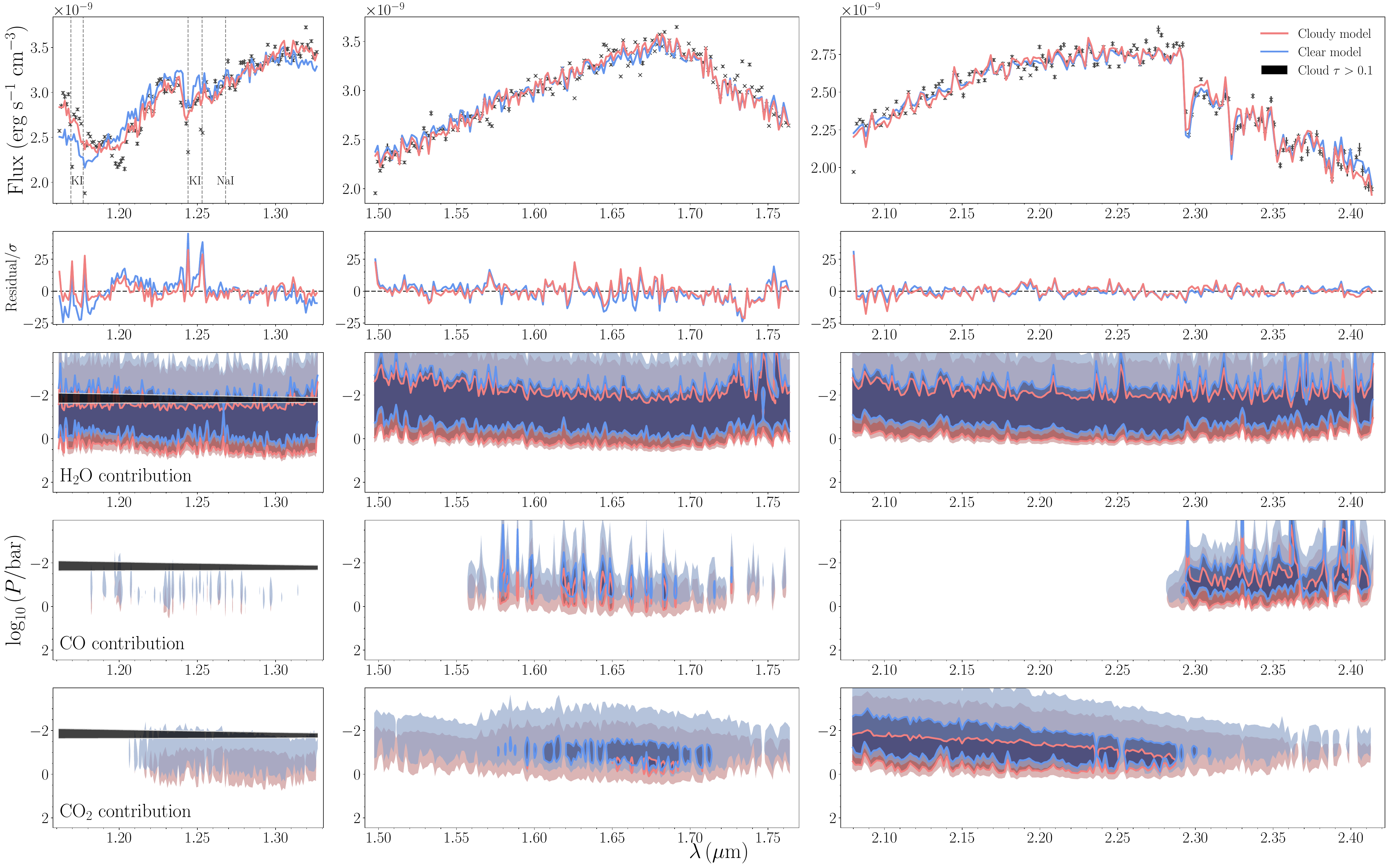}
\caption{Spectra and contribution functions of the retrieved forward model fits to the HD 106906 b data. The forward model spectra (using the MLE parameter values) are in color, with the results from the cloudy model in red and those of a cloud-free model in blue. The data are shown in grey. The retrievals are performed with the data binned down to a resolution 4--8 times lower than its original, to mitigate the potential effects of binning errors from the opacity tables. Prominent alkali lines in the $J$ band data are marked with vertical dashed lines, including 2 KI doublets and a smaller NaI line. Immediately beneath the spectra are the contribution functions for the 3 principal carbon and oxygen bearing species. The deepest contours, outlined in solid colors, enclose the regions where the contribution function reaches $>1$\% of the total contribution within the atmospheric column at a given wavelength bin. Each successive contour denotes 2 orders of magnitude smaller fractional contribution (here, $10^{-4}$).}
\label{fig:HD106906b_PLO_spectrum}
\end{center}
\end{figure*}

\begin{figure*}[htb]
\begin{center}
\textbf{Retrieval on HD 106906 b Spectrum} \\
\includegraphics[width=13cm]{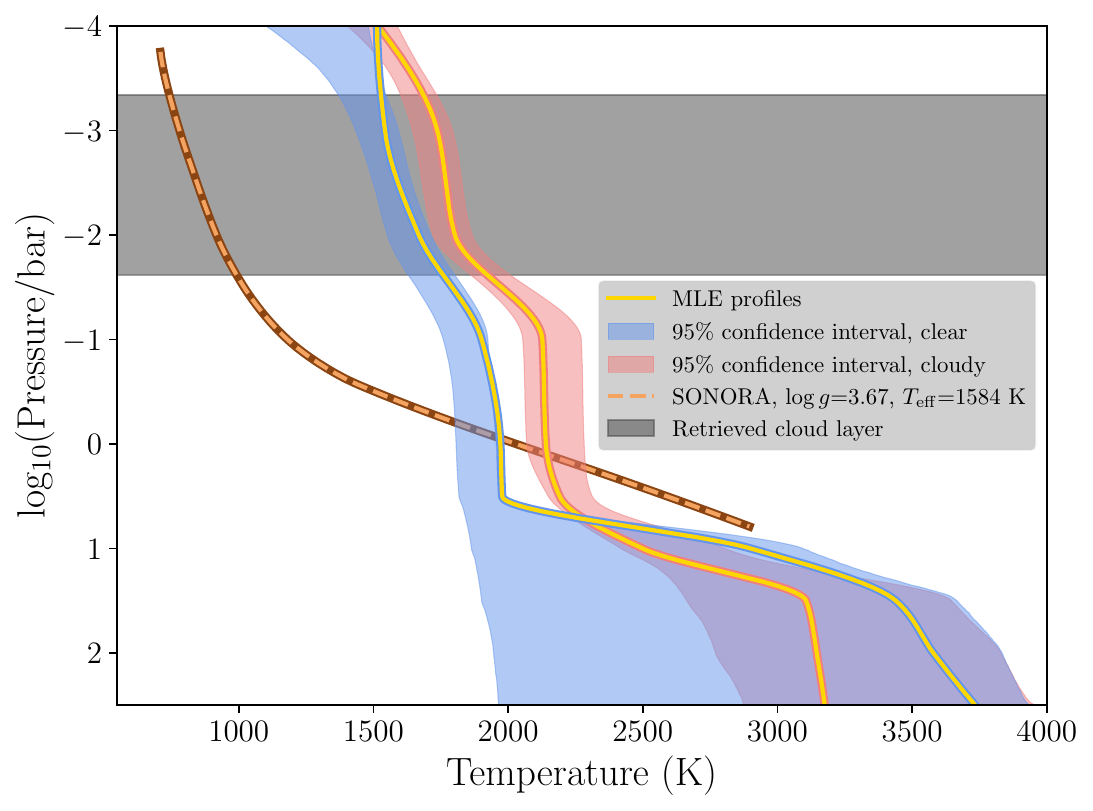}
\caption{The vertical temperature-pressure profiles of the retrieved forward model fits to the HD 106906 b data. We show the MLE, median, and 95\% confidence interval of the retrieved T-P profiles for the cloud-free (blue) and cloudy (red) models, the latter of which is described in \S \ref{sec:model:clouds}. Also plotted is the (cloud-free) \texttt{SONORA} model for a brown dwarf at the gravity and effective temperature of our best-fit cloudy model, with the change from radiative to convective behavior occurring at a few tenths of a bar. Our retrieved profiles in contrast vary much less in temperature with pressure down to the expected radiative-convective boundary for an object at this effective temperature.}
\label{fig:HD106906b_PLO_T-P}
\end{center}
\end{figure*}

\begin{figure*}[htb]
\begin{center}
\textbf{Retrieval on HD 106906 b Spectrum} \\
\includegraphics[width=17cm]{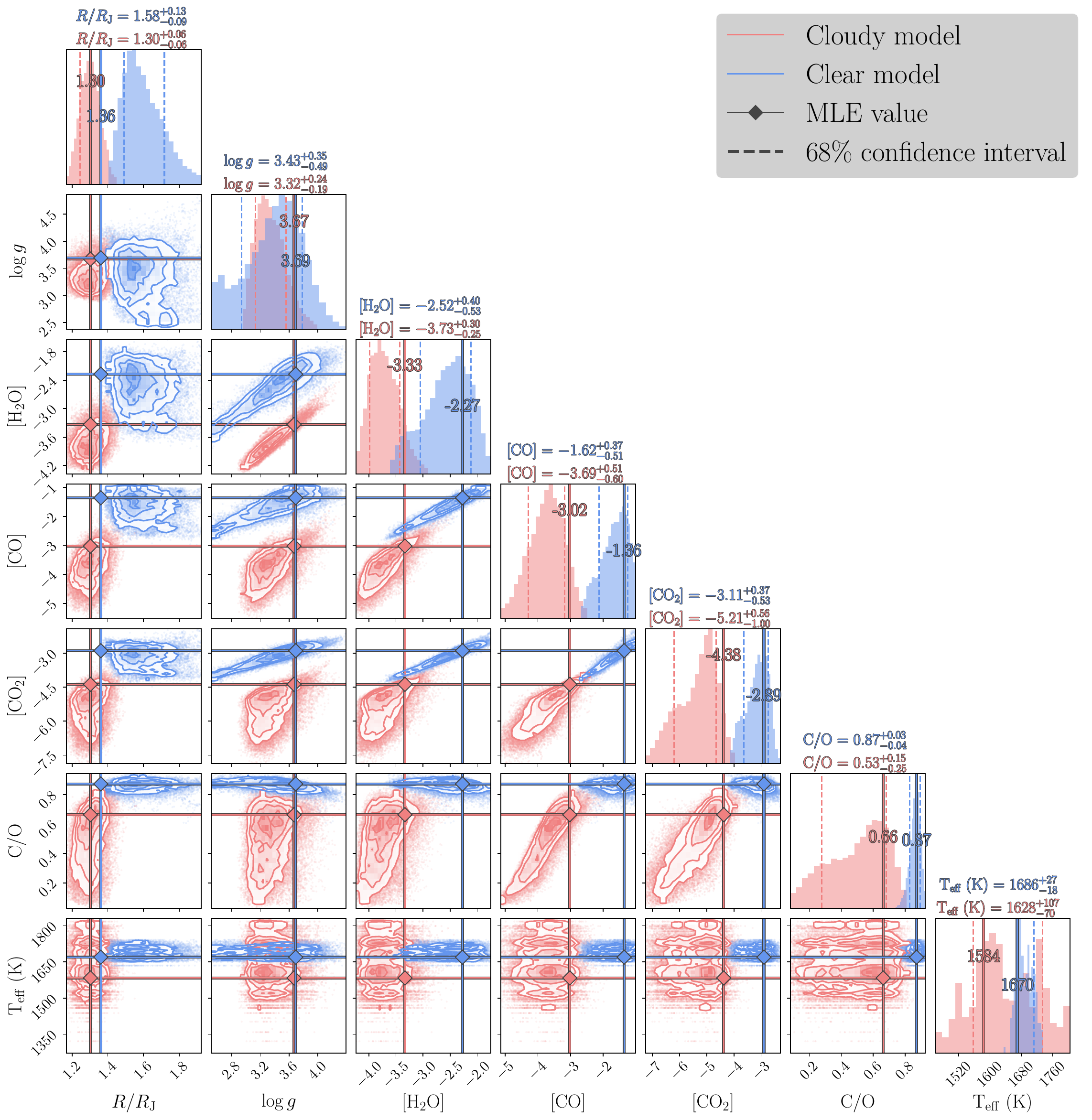}
\caption{A selection of parameters of the retrieved forward model fits to the HD 106906 b data, shown as 1-D and 2-D histograms in a corner plot of the retrieved posterior distributions for selected parameters. The median value and 68\% confidence interval of each parameter are shown at the top of each column; the full list of median, interval range, and MLE values are shown in Table \ref{table:HD106906b_parameters}.}
\label{fig:HD106906b_PLO_custom-corners}
\end{center}
\end{figure*}

\begin{figure*}[htb]
\begin{center}
\includegraphics[width=17cm]{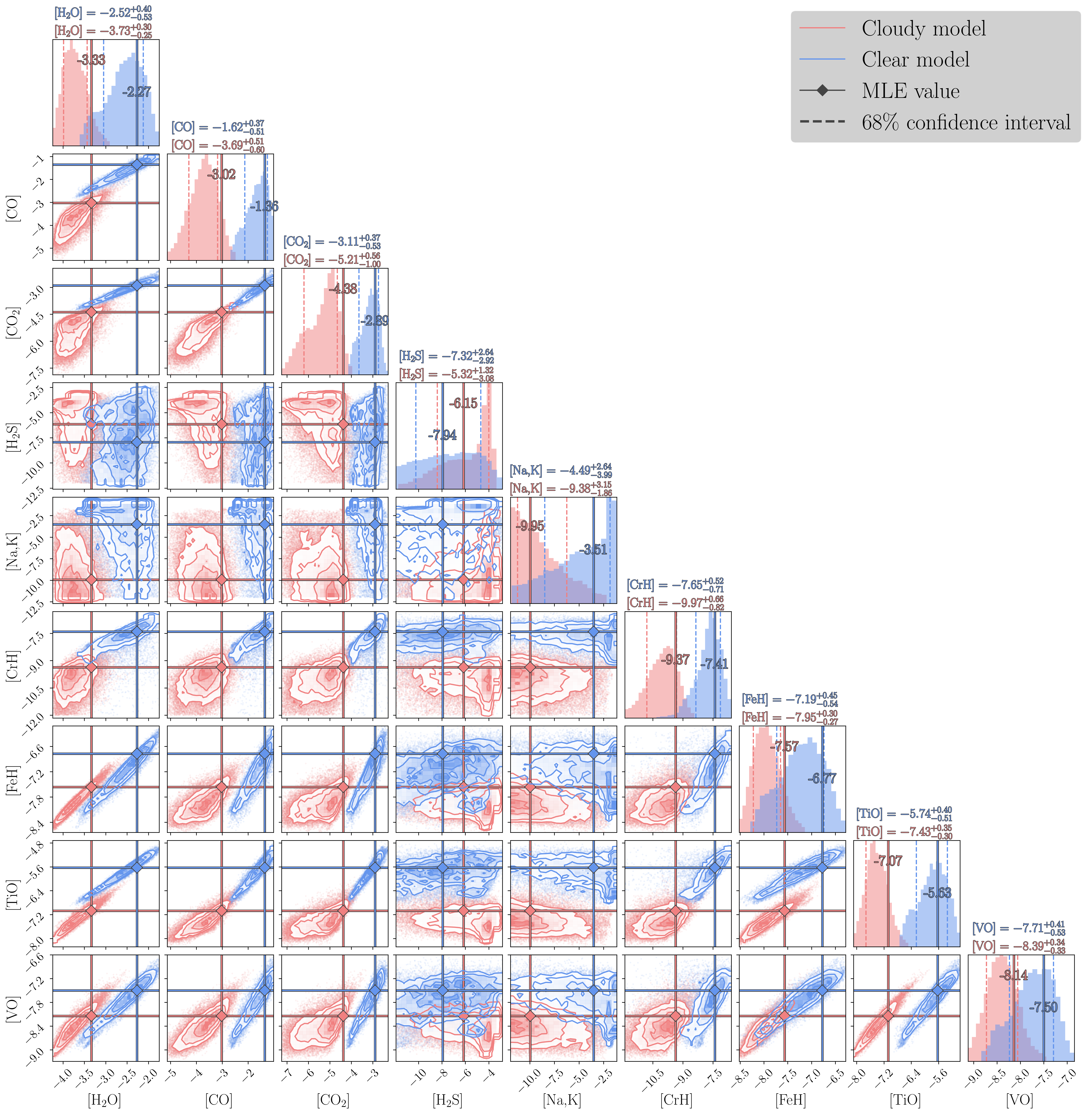}
\caption{Single-parameter (1-D) and parameter-versus-parameter (2-D) posterior distributions of gas parameters from the samples in the cloudy (red) and cloud-free (blue) forward model fit to the HD 106906b data. The cloud opacity is modeled as a power law in wavelength, as described in \S \ref{sec:model:clouds}. The median value and 68\% confidence interval of each parameter are shown at the top of each column; the full list of median, interval range, and MLE values are shown in Table \ref{table:HD106906b_parameters}.}
\label{fig:HD106906b_PLO_gas-corners}
\end{center}
\end{figure*}

\begin{figure*}[htb]
\begin{center}
\includegraphics[width=17cm]{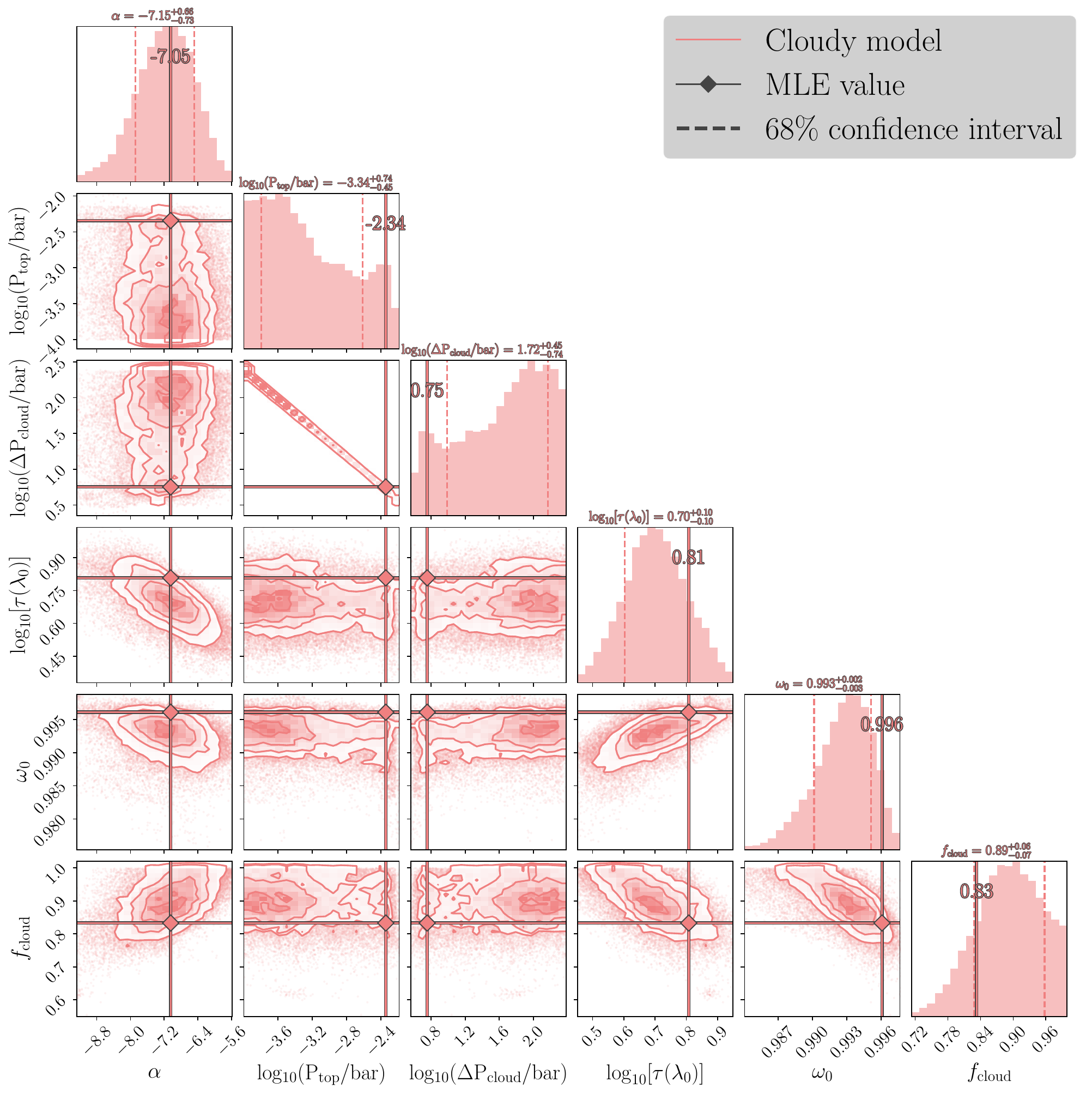}
\caption{Single-parameter (1-D) and parameter-versus-parameter (2-D) posterior distributions of cloud parameters from the samples in the cloudy forward model fit to the HD 106906b data. The cloud opacity is modeled as a power law in wavelength, as described in \S \ref{sec:model:clouds}. The median value and 68\% confidence interval of each parameter are shown at the top of each column; the full list of median, interval range, and MLE values are shown in Table \ref{table:HD106906b_parameters}.}
\label{fig:HD106906b_PLO_cloud-corners}
\end{center}
\end{figure*}

\begin{deluxetable}{lcccc}
\tabletypesize{\scriptsize}
\tablewidth{0pt}
\tablecaption{Median and MLE parameter values for the four model configurations used to retrieve on the spectra of HD 106906 b. \label{table:best-fit_parameters}}
\tablehead{ &
           \multicolumn{2}{c}{Cloudy Model} &
           \multicolumn{2}{c}{Cloud-free Model} \\
           \colhead{Name} &
           \colhead{Median} &
           \colhead{MLE} &
           \colhead{Median} &
           \colhead{MLE}
           }
\startdata
\cutinhead{Fit Quality}
$\chi_\nu^2$ & & 40.3 & & 60.5 \\
$\Delta \log\!\left(\mathrm{Bayes\,Factor}\right)$ & \multicolumn{2}{c}{56} & \multicolumn{2}{c}{$\left(0\right)$} \\
\cutinhead{Fundamental}
$R/R_\mathrm{J}$                                              & $1.43\pm0.05$ & 1.45 & $1.74\pm0.06$ & 1.66 \\
$\log_{10}\!\left[g/\!\left(\mathrm{cm~s}^{-2}\right)\right]$ & $3.32^{+0.24}_{-0.19}$ & 3.67 & $3.43^{+0.35}_{-0.49}$ & 3.69 \\
$M/M_\mathrm{J}$                                              & $1.92^{+1.48}_{-0.70}$ & 4.41 & $3.55^{+4.55}_{-2.39}$ & 6.62 \\
$T_\mathrm{eff}$ (K)                                          & $1628^{+107}_{-70}$ & 1584 & $1686^{+27}_{-18}$ & 1670 \\
C/O                                                           & $0.53^{+0.15}_{-0.25}$ & 0.66 & $0.87^{+0.03}_{-0.04}$ & 0.87 \\
Metallicity                                                   & $-0.24^{+0.41}_{-0.35}$ & 0.26 & $1.66^{+0.34}_{-0.38}$ & 1.85 \\
\cutinhead{Gases ($\log_{10}$ number abundance)}
H$_2$O   & $-3.73^{+0.30}_{-0.25}$ & $-3.33$ & $-2.52^{+0.40}_{-0.53}$ & $-2.27$ \\
CO       & $-3.69^{+0.51}_{-0.60}$ & $-3.02$ & $-1.62^{+0.37}_{-0.51}$ & $-1.36$ \\
CO$_2$   & $-5.21^{+0.56}_{-1.00}$ & $-4.38$ & $-3.11^{+0.37}_{-0.53}$ & $-2.89$ \\
H$_2$S   & $-5.32^{+1.32}_{-3.08}$ & $-6.15$ & $-7.32^{+2.64}_{-2.92}$ & $-7.94$ \\
Na+K     & $-9.38^{+3.15}_{-1.86}$ & $-9.95$ & $-4.49^{+2.64}_{-3.99}$ & $-3.51$ \\
CrH      & $-9.97^{+0.66}_{-0.82}$ & $-9.37$ & $-7.65^{+0.52}_{-0.71}$ & $-7.41$ \\
FeH      & $-7.95^{+0.30}_{-0.27}$ & $-7.57$ & $-7.19^{+0.45}_{-0.54}$ & $-6.77$ \\
TiO      & $-7.43^{+0.35}_{-0.30}$ & $-7.07$ & $-5.74^{+0.40}_{-0.51}$ & $-5.63$ \\
VO       & $-8.39^{+0.34}_{-0.33}$ & $-8.14$ & $-7.71^{+0.41}_{-0.53}$ & $-7.50$ \\
\cutinhead{Temperature-Pressure}
$T_{-4}$  (K)  & $1499^{+48}_{-53}$   & 1516 & $1349^{+80}_{-115}$   & 1514 \\
$T_{-3}$  (K)  & $1720^{+33}_{-36}$   & 1737 & $1525^{+37}_{-43}$    & 1543 \\
$T_{-2}$  (K)  & $1791^{+32}_{-35}$   & 1802 & $1638^{+42}_{-47}$    & 1668 \\
$T_{-1}$  (K)  & $2156^{+51}_{-54}$   & 2127 & $1848^{+44}_{-46}$    & 1902 \\
$T_{0}$   (K)  & $2172^{+50}_{-54}$   & 2140 & $1874\pm39$           & 1971 \\
$T_{0.5}$ (K)  & $2233^{+39}_{-42}$   & 2191 & $1887^{+39}_{-42}$    & 1978 \\
$T_{1}$   (K)  & $2618^{+102}_{-100}$ & 2499 & $2041^{+429}_{-133}$  & 2901 \\
$T_{1.5}$ (K)  & $3120^{+304}_{-253}$ & 3107 & $2373^{+687}_{-380}$  & 3440 \\
$T_{2}$   (K)  & $3270^{+344}_{-289}$ & 3146 & $2728^{+697}_{-566}$  & 3579 \\
$T_{2.5}$ (K)  & $3468^{+330}_{-319}$ & 3176 & $2988^{+677}_{-742}$  & 3735 \\
\cutinhead{Clouds}
$\alpha$                                                         & $-7.15^{+0.66}_{-0.73}$   & $-7.05$ & & \\
$\log_{10}\!\left( P_\mathrm{top}/\mathrm{bar} \right)$          & $-3.34^{+0.74}_{-0.45}$   & $-2.34$ & & \\
$\log_{10}\!\left( \Delta P_\mathrm{cloud}/\mathrm{bar} \right)$ & $1.72^{+0.45}_{-0.74}$    & $0.75$  & & \\
$\log_{10}\!\left[\tau\!\left(\lambda_0\right)\right]$           & $0.70\pm0.10$             & 0.81    & & \\
$\omega_0$                                                       & $0.993^{+0.002}_{-0.003}$ & 0.996   & & \\
$f_\mathrm{cloud}$                                               & $0.89^{+0.06}_{-0.07}$    & 0.83    & & \\
\cutinhead{Photometric Calibration}
Calibration factor $\left(J\right)$ & $1.05\pm0.04$ & 1.07 & $0.97^{+0.05}_{-0.04}$ & 0.89 \\
Calibration factor $\left(H\right)$ & $0.88\pm0.02$ & 0.89 & $0.96\pm0.02$          & 0.93 \\
\enddata
\end{deluxetable}

Results from the retrieval with our cloud model included are shown in Figure \ref{fig:HD106906b_PLO_spectrum} for the spectra and contribution functions, Figure \ref{fig:HD106906b_PLO_T-P} for the retrieved T-P profiles, and posterior distributions of parameters in Figures \ref{fig:HD106906b_PLO_custom-corners}--\ref{fig:HD106906b_PLO_cloud-corners}. The retrieved model captures much of the broad shape of the spectrum, but fails to capture the amplitudes of some of the features in the $J$ band. Given the S/N of the data, the $\chi_\nu^2$ statistic of the model indicates a poor fit to the data, at $\chi^2_\nu\approx 40$. As noted in \S \ref{sec:results:setup}, we have scaled the original uncertainties assuming the errors are uncorrelated; this statistic assumes the most optimistic noise model and therefore the most pessimistic fit quality. The retrieved radius posterior range ($1.30\pm0.06 \, R_\mathrm{J}$) is smaller than the radius one derives from the best-fit bolometric luminosity and effective temperature in \citet{Daemgen2017}, which is 1.47 $R_\mathrm{J}$. However, this is affected by our choice to normalize the spectrum such that the $K_\mathrm{s}$ band calibration factor is 1; therefore, there is a bit of ambiguity in whether our retrieved radius is strictly consistent or inconsistent with these previous constraints. Our inferred effective temperature is high given the small radius. The retrieved surface gravity is low compared with that derived from the \citet{Daemgen2017} fundamental parameters ($\log g = 4.19\pm0.40$), though the 95\% confidence interval does overlap with this range. This low gravity, combined with the radius, yields a 68\% confidence interval of $1.92^{+1.48}_{-0.70}$ $M_\mathrm{J}$ for the mass, with an MLE value of 4.41 $M_\mathrm{J}$. This is smaller than the mass range of $11\pm2$ $M_\mathrm{J}$ from evolutionary models as presented in \citet{Bailey2014}, as well as the estimated 13 $M_\mathrm{J}$ mass if one were to adopt the mean age of the LCC, at 17 Myr. We are unlikely to be able to disentangle the low retrieved mass from the existing degeneracies that persist between gravity, metallicity, and the T-P profile. Additionally, the range of bolometric luminosities we infer from our results is low compared with the original evolutionary model constraints: our cloudy model returns a 68\% confidence interval of $\log_{10}\!\left(L/L_\odot\right) = -3.94\pm0.10$, while the cloud-free model returns $\log_{10}\!\left(L/L_\odot\right) = -3.73_{-0.06}^{+0.07}$, compared with the original constraint of $\log_{10}\!\left(L/L_\odot\right) = -3.64\pm0.08$. Only the cloud-free model is consistent with the \citet{Bailey2014} constraints. The primary reason for the disagreement between the original luminosity estimates and those from our cloudy model is due to the fact that our effective temperature is computed by generating a low-resolution forward model over a longer wavelength range than the data, at 0.6--30 $\mu$m. With such a large negative exponent to the wavelength dependence of our retrieved cloud model, the emission bluer than the data wavelength will be suppressed, cooling the derived effective temperature. However, our cloudy model is consistent with the luminosity ranges derived using subsets of ``young'' (``YNG'' and ``YNG2'') targets, as presented in Table 19 of \citet{Faherty2016}; \citet{Daemgen2017} used this to calculate a luminosity constraint of $\log_{10}\!\left(L/L_\odot\right) = -3.83\pm0.35$ and $-3.64\pm0.24$ for the YNG and YNG2 relations, respectively.

Our retrieved C/O ratio of $0.53^{+0.15}_{-0.25}$ is consistent with the estimated C/O ratio distribution of the stellar association in which HD 106906 resides ($0.52\pm0.11$; see Equation \ref{eq:ctoo_Sco-Cen}); the 3 primary C+O constituents (H$_2$O, CO, and CO$_2$) are constrained to within 0.5--1 dex and show positive correlations among each other, as well as with the surface gravity. The correlation between molecular abundances and gravity is known to be a consequence of a degeneracy where, in fitting absorption features, the flattening effect of higher gravity can be at least partially offset by higher abundances \citep[see e.g.][]{Todorov2016}. The full posterior distributions for gas abundances are shown in Figure \ref{fig:HD106906b_PLO_gas-corners}. The retrieved H$_2$O abundance for the best-fit model ($-3.33$ dex) is within 0.1 dex of the expectation given the retrieved T-P profile, if one assumes chemical equilibrium for an object at solar metallicity and C/O ratio ($-3.35$ dex). The CO abundance, at $-3.02$ dex, is higher than the equilibrium value of $-3.28$, which drives the best-fit C/O to just beyond the 68\% confidence interval, at 0.66. CO's impact is comparable to that of H$_2$O but over isolated regions of the spectrum; the bulk metallicity has uncertainties of order 0.4 dex but is consistent with solar metallicity as well as the metallicity range of its stellar association. The CO$_2$ abundance ($-5.21^{+0.56}_{-1.00}$ dex, best-fit value $-4.38$ dex) is the least constrained of the 3 major C+O molecular absorbers, and has the smallest effect on the C/O ratio. The absorbers least consistent with an equilibrium abundance are the alkalis; with a range of $-9.38^{+3.15}_{-1.86}$ and a best-fit abundance of $-9.97$ dex, the model essentially ignores the alkali absorption features in its fit. This is surprising since there are prominent absorption lines of potassium (two KI doublets) in the $J$ band. The failure of the model to capture these absorption lines appears to be a consequence of reducing the spectral resolution; previous attempts at retrievals were less successful at converging to a global atmospheric fit than the ones shown in this work, but had enough of the line shape at the original resolution to fit the abundances, as well as using the ratio of the KI doublet line depths as an additional constraint on gravity. Such models return alkali log abundances ranging from about $-5$ to $-7$, but also return infeasibly high gravities, exceeding $\log g = 6$ unless a restrictive prior is used.

The preferred T-P profile is shallow in its temperature gradient from the top of the atmosphere to a pressure of several bars, after which point the temperature rapidly increases to approximately 3100 K at several tens of bars. The profile then returns to a nearly isothermal behavior to the base of the model atmosphere. Figure \ref{fig:HD106906b_PLO_T-P} shows the profile along with its cloud-free counterpart and a cloud-free \texttt{SONORA} brown dwarf profile interpolated to match the maximum-likelihood gravity and effective temperature from the cloudy model. In contrast with the radiative-convective equilibrium profile from \texttt{SONORA}, with a shallow thermal gradient gradually increasing to a higher adiabatic gradient at the radiative-convective boundary, our retrieved profile can be described as nearly isothermal layers for the log-pressure ranges of $\sim -3$ to $-2$, again from $\sim -1$ to 0.5, and, at least for the cloudy model, a nearly isothermal layer at the deepest $\sim 1$ dex of the model pressure range. These nearly isothermal ``layers'' are punctuated with comparatively rapid temperature increases. The majority of the contribution from the major absorbers (H$_2$O, CO, and CO$_2$) comes from pressures of $\sim 1$ mbar--1 bar. The full posterior distributions for the cloud parameters are shown in Figure \ref{fig:HD106906b_PLO_cloud-corners}. The distribution of cloud top pressures ranges from the very top of the model ($10^{-4}$ bar) to a few mbar, and the top pressure is strongly correlated with the depth of the cloud. The maximum pressure of the cloud appears to be the most important parameter here, which when combined with the fact that most of the gas contribution to the emission is beneath this maximum pressure, implies that the model prefers whichever cloud layer can produce some fixed total column optical depth. With a highly negative power-law exponent ($\alpha=-7.15^{+0.66}_{-0.73}$, best-fit value $-7.05$), clouds produce significant opacity only for the $J$ band. As with the retrieved clouds for 2M2224, this might not reflect an accurate constraint on actual cloud opacity, but for the model is a way to suppress emission in the bluest wavelengths without an obvious physical interpretation. The retrieved distribution of the single-scattering albedo $\omega_0$ is tightly distributed and is close to the upper limit. The covering fraction ($f_\mathrm{cloud}$) distribution is consistent with but not centered at 1, which corresponds to a near global coverage of a very reflective cloud.

\subsection{The Cloud-free Model}\label{sec:results:cloud-free}
The spectral fits, contributions, profiles, and posterior distributions are plotted with the cloudy solution in Figures \ref{fig:HD106906b_PLO_spectrum}--\ref{fig:HD106906b_PLO_cloud-corners}. Excluding clouds entirely in our model returns a worse fit quality, with $\chi_\nu^2 = 61$. The log-ratio of the Bayes factors is 56, indicating the cloudy solution provides overwhelmingly stronger evidence than the cloud-free solution. The main reduction in fit quality is in the $J$ band, consistent with where the cloudy solution places most of its cloudy opacity. While the gravity ($\log g = 3.43^{+0.35}_{-0.49}$, best-fit value 3.69) is consistent with the cloudy fit, the radius of the cloud-free solution increases to $1.74\pm0.06$ $R_\mathrm{J}$, and the abundances of all species except H$_2$S increase by amounts ranging from 0.5--2 dex. The degeneracy between gravity and the molecular abundances is even stronger than that of the cloudy atmospheres, with the CO abundance rising more than the H$_2$O abundance, yielding a C/O distribution of $0.87^{+0.03}_{-0.04}$. Since the H$_2$O contribution dominates in the $J$ band, it is possible that the relatively poor fit of the cloud-free model in the $J$ band affects the accuracy of the retrieved C/O ratio. The T-P profile shows a similar shape to that of the cloudy profile, albeit shifted by roughly 100--200 K from the model top to a pressure of a few bars, deeper than which the gradient increases in a similar fashion to the cloudy profile, but with a much higher uncertainty that is also consistent with an isothermal profile.

\section{Discussion}\label{sec:discussion}
In the comparison of our results from \S \ref{sec:test-results} concerning 2M2224 with those of \citet{Burningham2017}, we retrieve a C/O ratio whose confidence interval overlaps that of their posterior distributions --- but our model disagrees on nearly everything else, including the radius, gravity, T-P profile, cloud properties, and absolute abundances. In the case of 2M2224, we have the hindsight provided by \citet{Burningham2021} that, with mid-infrared data, they were able to distinguish between specific cloud particle compositions. As a result, the slope of their T-P profile decreased considerably, as well as the retrieved abundances of H$_2$O and CO by roughly 0.4 dex. Therefore, it is not surprising that different retrieval codes disagree when the available wavelength ranges differ. However, the fact that our codes nevertheless converged on the same C/O ratio is promising. A recent analysis from \citet{Rowland2023} shows that in the L-dwarf regime, and especially earlier L dwarfs, the choice of T-P parametrization (or non-parametrization) matters in the near-infrared: more restrictive or smoothed parametrizations may bias the retrieved parameters. Therefore, the difference between our retrieved atmospheric profiles and abundances and those of \citet{Burningham2017} may lie primarily in our differing choices of T-P parametrization. Additionally, both of the retrievals on 2M2224 assume a constant abundance with pressure for all species; \citet{Rowland2023} find that one must account for non-uniform abundances in FeH in particular due to rainout chemistry in order to avoid biasing the retrieved T-P profile. This effect is strongest for early L dwarfs, and may not bias the existing results as severely for 2M2224 at a spectral type of L4.5. However, it suggests we may consider non-uniform chemistry in future retrievals for objects such as HD 106906 b, which lies at L0.5. We discuss our current retrieval results below.

Our retrieved C/O ratio of $0.53^{+0.15}_{-0.25}$ for HD 106906 b is entirely consistent with our estimate for the C/O ratios of fellow members of the Sco-Cen association ($0.52\pm0.11$, as in \S \ref{sec:data:data}). Therefore, our results with the cloudy model do not rule out a stellar-like, brown dwarf companion formation pathway for HD 106906 b. However, our model returns a T-P profile whose shape is unlikely to be entirely physical, alternating between regions of nearly isothermal behavior with regions of rapid temperature increases. Unlike the shallow wavelength dependence of the cloud opacity in our simulated models, the cloudy fits on the HD 106906 b data show a large negative exponent, indicating clouds contribute primarily in the $J$ band, but relatively little at redder wavelengths. This means that the differences matter more in the retrieved profiles between the cloud-free and cloudy models; both fit the $H$ and $K_\mathrm{s}$ band spectra with similar quality, but the cloudy model adjusts the absolute abundances and temperatures in accordance with the cloud constraints from the $J$ band, ``breaking'' the degeneracy between the T-P profile, gravity, and absolute abundances. In the retrieval of 2M2224's atmosphere, we stopped short of invoking the interpretation of \citet{Tremblin2015} to characterize the shallow thermal gradients, as our models retrieved significant cloud opacity across the wavelength range of our data. Here in contrast the cloudy model prefers little to no cloud opacity in $H$ and $K_\mathrm{s}$, which keeps viable the interpretation of the data as representing a thermo-chemical instability driven by dis-equilibrium chemistry. However, the confidence in any claim, whether in the accuracy of absolute molecular abundances or characterizing the vertical structure, is limited in the absence of wider spectral wavelength coverage that can increase the precision of the retrieved profile and also capture key signatures of cloud condensate species.

As mentioned above, the retrieved cloud opacity is heavily biased toward shorter wavelengths, with the cloud opacity only reaching an optical depth of $\sim 1$ in the $J$ band. Such a strong wavelength dependence, with a power-law exponent of $-7$, is likely not to be attributable to a specific condensate in the atmosphere, and may be a combination of some cloud opacity (e.g.~SiO$_2$, as seen in the constraints in the condensate pressures of \citealt{Burningham2021}) and potential remaining systematics in the shortest wavelengths. This potential degeneracy is likely only resolved with broader wavelength coverage, which is proving increasingly invaluable for accurate atmospheric characterization, and/or a more sophisticated treatment of clouds, such as modeling multiple distinct cloud layers. An additional drawback to modeling clouds using a functional form for the opacity, rather than incorporating scattering from model cloud particles, is that we are not able to account for any amount of carbon and oxygen contained within the clouds. \citet{Burningham2021} were able to determine that the choice of cloud model has an effect on their C/O ratio constraints at only around the 1\% level, which means the C/O ratios are consistent within their retrieved uncertainties. In their case it was primarily because they found their oxygen-bearing clouds to reside primarily at pressures shallower than the photosphere, meaning their contribution to the overall oxygen budget was $\sim$1\%. We can make a first-order estimate of the maximum effect of silicate condensation on our C/O ratio by following the prescription of \citet{Burrows1999}, used in \citet{Line2015,Line2017,Burningham2021}, that assumes on average 3.28 atoms of oxygen are sequestered per silicon atom in silicate condensates. Since our retrieved metallicity distribution is consistent with solar, as are our abundance distributions for the major oxygen-containing species when compared with solar-metallicity equilibrium models, we can estimate that a maximum of $\sim 16$\% of our atmospheric oxygen may be held in silicate clouds. Our best-fit C/O ratios would then drop to as low as 0.55 (vs. 0.66), with the retrieved range updated to $0.45^{+0.13}_{-0.21}$ (vs. $0.53^{+0.15}_{-0.25}$), still consistent with the association C/O. However, a true accounting of the oxygen budget in condensates will necessitate a more careful treatment of clouds than this work provides.

The limitations of the near-infrared in characterizing atmospheres in this temperature range are now well-established. Therefore, our work serves as a preparation for future retrievals, taking advantage of a broader wavelength coverage, on this and other similar planetary-mass companions. \JWST{} now allows high-resolution, high signal-to-noise emission spectra of spatially resolved, very low-mass companions, and the largest benefit to retrievals is its ability to extend to the mid-infrared. In the case of HD 106906 b, GTO observations with \JWST{} are already scheduled that will capture both an $R\sim$ 1000 spectrum using  NIRSpec (G395M, $\lambda=2.87$--5.27 $\mu$m) and a low resolution ($R\sim$ 100) MIRI LRS spectrum spanning 5--12 $\mu$m. The results of \citet{Burningham2021} have suggested that extending into the mid-infrared not only constrains specific cloud compositions, but also significantly increases the range of the spectrum little affected by cloud opacity, which can allow for more accurate constraints on both the T-P profile and gas opacities. The results for 2M2224 suggests that the relative gas abundances may be robust to limitations in wavelength range, but that one should not expect consistent gravity, T-P profile, or cloud constraints unless one has longer wavelength coverage. This being said, we are still limited to regions of the atmosphere that can be seen in emission; longer wavelengths will tend to probe cooler regions, which for directly imaged companions without thermal inversions will mean shallower pressures. The deepest parts of the atmosphere beneath the photosphere for these wavelengths, and/or beneath optically thick cloud layers, may still be inaccessible. This means that a complete picture of metrics such as the C/O ratio are likely to be still out of reach.

\section{Conclusions}\label{sec:conclusion}
We use an atmospheric retrieval code, the \APOLLO{} code, in its first application to a cloudy L dwarf spectrum. Our goal is to constrain formation pathways of companions to young stars. Signatures of their formation as either binary star-like or planet-like (i.e.~formed in a disk) should be imprinted in their chemistry, using metrics such as the C/O ratio and metallicity.

From the analysis of our model results, we conclude that:
\begin{itemize}
    \item Based on our self-retrieval results, the wavelength range and signal-to-noise of the HD 106906 b is sufficient to accurately constrain the C/O ratio for simulated data. Cloud-free models can retrieve a similarly accurate C/O ratio but are not preferred statistically when cloud opacity is present in the data, and in that case may return inaccurate gravities, radii, and particularly bias toward high molecular abundances.
    \item When comparing our retrieval results on the field L dwarf 2M2224-0158 with those of \citet{Burningham2017}, we find a consistency in our C/O ratios but a disagreement in the T-P profiles, cloud properties, and molecular abundances. This warrants a similar interpretation to those in \citet{Burningham2017} and \citet{Molliere2020} where a degeneracy is seen, especially in the near-infrared, between retrieved cloud properties and the T-P profile.
    \item Our best-fitting model for HD 106906 b yields a C/O ratio of $0.53^{+0.15}_{-0.25}$, consistent with the range of C/O ratios estimated for members of the Sco-Cen association ($0.52\pm0.11$). This implies that we cannot rule out the hypothesis that HD 106906 b formed via the pathway expected for a brown dwarf companion to HD 106906, in contrast with a planet-like pathway.
    \item However, our solution for the atmospheric emission of HD 106906 b yields negligible cloud opacity in the $H$ and $K_\mathrm{s}$ bands, which along with a shallow temperature gradient at pressures less than a few bars, suggest that our results point to a cloud-temperature degeneracy.
    \item As with many other retrievals of objects at similar masses and temperatures, additional data in the mid-infrared ($\gtrsim 10$ $\mu$m) will be helpful in breaking the degeneracies in atmospheric structure and composition.
\end{itemize}
\JWST{} is currently observing directly imaged companions, obtaining spectra at resolutions $\gtrsim 1000$ at wavelength ranges not obtainable from the ground. By expanding the region where clouds are expected to contribute little, such as the thermal infrared region of $\sim 3$--5 $\mu$m, we can better constrain the thermal structure and gas abundances, and by extension both the gravity and metallicity. Additionally, $R \sim 100$ spectra are available through the mid-infrared instrument (MIRI), which extends the wavelength range into the realm where cloud-specific features --- such as those from enstatite --- are visible in emission. Follow-up observations in these wavelength ranges are planned for HD 106906 b that will allow us to employ a cloud model that more directly models specific cloud condensates. Additionally, while we have not resolved the discrepancy in brightnesses in the near-infrared between ground- and space-based observations, additional wavelength coverage into the thermal and mid-infrared with \JWST{} will also help us investigate this apparent disagreement. At the same time, accurate C/O ratios and metallicities of more companion hosts are needed to directly compare with the retrieved chemistry of the companions. In either case, retrievals on either ground-based or space-based data will benefit greatly from a set of standardized inter-model comparisons of results from various retrieval codes, to test how each model's treatment of the physics affects the inferred atmospheric properties.

\acknowledgments
We would like to acknowledge Dr. Natasha Batalha, whose advice and expertise on retrieval and statistical methods has been valuable in improving the rigor of this work. We also would like to thank the careful thought and feedback of our referees.

We are grateful for support from NASA through the \JWST{} NIRCam project though contract number NAS5-02105 (M. Rieke, University of Arizona, PI).

\vspace{5mm}

\software{APOLLO \citep{Howe2017,Howe2022}, Astropy \citep{ast13}, Jupyter \citep{klu16}, Matplotlib \citep{hun07}, Numpy \citep{van11}, Pandas \citep{mck10}, PICASO \citep{Batalha2019}, Scipy \citep{jon01}, Synphot \citep{pysynphot,synphot}}

\bibliography{library}{}
\bibliographystyle{aasjournal}

\end{document}